# Quantifying how AI Panels improve precision

Nicholas Beale MA FIMA Sciteb 20 April 2026

Nicholas.beale@sciteb.com

## Abstract

AI in applications like screening job applicants had become widespread, and may contribute to unemployment especially among the young. Biases in the AIs may become baked into the job selection process, but even in their absence, reliance on a single AI is problematic. In this paper we derive a simple formula to estimate, or at least place an upper bound on, the precision of such approaches for data resembling realistic CVs:

$$\mathrm{P(q)} \approx \frac{\rho \mathrm{n}^b + \mathrm{q}(1-\rho)}{1+(\mathrm{n}^b - 1)\rho} \quad \text{where } b \approx q^* + 0.8\,(1-\rho) \text{ and } q^* \text{ is q clipped to } [0.07, 0.22] \qquad \text{(Equation 1)}$$

where P(q) is the precision of the top q quantile selected by a panel of n AIs and ρ is their average pairwise correlation. This equation provides a basis for considering how many AIs should be used in a Panel, depending on the importance of the decision. A quantitative discussion of the merits of using a diverse panel of AIs to support decision-making in such areas will move away from dangerous reliance on single AI systems and encourage a balanced assessment of the extent to which diversity needs to be built into the AI parts of the socioeconomic systems that are so important for our future.

## Introduction

Using AI to support human judgement is increasingly pervasive. One example is the use of AI to screen CVs – a seemingly mundane task which has huge implications for the development of any organisation – and to fix intuition this is the example that will be used in this paper.

Over-reliance on AI is dangerous, and it is especially dangerous to rely on a single AI system. The aim of this paper is to provide a straightforward practical way in which practitioners can use multiple commercially available AIs and employ a theoretically sound and empirically tested equation to estimate how much extra precision they can expect from deploying additional AIs or, conversely, how many AIs they may need to deploy to achieve a desired precision.

CV ranking and similar judgement-oriented tasks are not like the problems with very specific solutions which have been a major focus of AI research, evaluation and benchmarking. Both have their place, but it is arguable that the pervasiveness of AIs in judgement-like tasks is at least as serious a socio-economic problem as AIs taking over specific solution-oriented functions.

We suggest that anyone using AI to help them make important decisions should ask:

i. How important is it to get this decision right?

And therefore:

ii. How much diversity of input do we need from various AIs (and other sources) to assist in this?

## Background

There is a widespread view around AI, at least in the West, that we are in a winner takes all race for a quasi-godlike superintelligence. It is interesting that a serious author [Mallaby 2026:p xiii] writes of "an omniscient machine…that would occupy, effectively, that position in the cosmos that religious believers once ascribed to an all-powerful divinity" apparently unaware of the logical [Godel 1931, Lucas 1961, Penrose 1989], and physical impossibility of even an omniscient machine, let alone an omnipotent one. But even the more aware proponents of this kind of view suggest that, when the winner develops this super-machine, it will self-improve exponentially and make all alternatives obsolete. An alternative viewpoint is that, whilst AI can certainly be valuable, it is inherently imperfect and unreliable. It is therefore wiser, if you are using AI to help make important decisions, to use multiple AIs in some form of a "Panel" and enable them to cross-check each other. This has been a fundamental principle of fault tolerant computing since at least [Avižienis, 1967] but hasn't been salient in AI discussions. Ensemble machine learning approaches have a long history, going back at least to [Schapire 1990] but the main focus has been on combining multiple instances of broadly the same algorithms. However, as [Hong & Page 2004] demonstrated formally, diverse problem-solving agents have significant advantages. And the specific case of heterogeneous AI panels, with diversity of architecture and training, has received much less attention or formal quantitative treatment.

[Lakshminarayanan et al. 2017] showed that multiple independently-trained networks improve calibration and uncertainty, but this is primarily focused on homogeneous networks, with essentially the same architecture. [Wang et al. 2024] introduces the notion of Mixtures of Agents with distinct LLMs, but the architecture they describe is very different and quite hard to implement, and although they have some nice empirical results there is no underlying formula. More recently {Li & al 2025] advocate a "self-MoA" approach and show that, in certain contexts, it can be more effective to obtain multiple outputs from the best model for a particular task.

In contrast to the work described above, the approach described in this paper is readily available to any organisation with access to a number of AIs, but without any special additional software. In addition, by providing a simple quantitative formula, organisations can examine the trade-offs between increased diversity and improved precision – although we would stress that the real imprecision of AI estimates is likely to be significantly higher than the calculations would suggest due to common assumptions across all current AI models.

## Mathematical Overview

To fix the intuition on this work we will discuss a situation where there are *m* candidates for a job and we have a CV (US: resumé) for each, and an AI is asked to grade the candidates on a scale of 0.0 to 10.0 and select the top q%. Straightforward extensions to other situations are left to the reader. It is worth noting formally that the use of AIs in this context is increasingly widespread and has been widely recognized as problematic. See eg [Raghavan et al 2020], [Chen 2023] and papers citing them.

Suppose we have *m* CVs whose "true" scores are $v_j$ and an AI estimates each score as $x_j$. For $q \in [0,1]$ define Top($q$,$x$) as the set of $qm$ CVs with the highest scores in x, and define the Precision P as:

$P(q; x, \nu) \coloneqq \frac{Card(Top(q,x) \cap Top(q,\nu))}{qm}$ with the *x* and *v* dropped when they are clear from context.

We are going to use multiple AIs so let us denote by X_ij the score that AI i assigns to CV j, and assume that:

$$X_{ij} = a_i \nu_j + c_i + \epsilon_{ij} \qquad (2)$$

Note that, although it is mathematically convenlient to consider the noise terms as independent, this will not be true in practice. Different AIs are trained on highly overlapping datasets, share underlying methodologies, and notoriously tend to embed social prejudices [Caliskan et al 2017] [Ferrara 2024]. So even if there is a high agreement among AIs on a rating this does not necessarily mean it's right.

We first observe with real data that, if we look at the average performances of single AIs whose average correlation coefficient is ρ:

$$P(q) \approx \rho + q\,(1 - \rho) \qquad (3)$$

This is related to a result of [Brogden 1946] and extends to larger panels with the appropriate $\rho_n$. We then observe that P(q) at any given level improves with n, but careful simulation shows that the classic [Brown 1910] scaling needs a correction parameter *b* (<1) and that the effective correlation coefficient $\rho_n$ of an independent panel of n AIs for this purpose is:

$$\rho_n \approx \frac{\rho n^b}{1+(n^b-1)\rho} \qquad (4)$$

With $b \approx q^* + 0.8\,(1 - \rho)$ and $q^*$ is q clipped to [0.07, 0.22] (4a)

*b* is found empirically, we have found no way to derive it. It seems not to matter much whether the data are normally distributed or whether, as in real datasets, they have somewhat fat tails. Algebra gives (1).

Note that this is an upper bound because it assumes the AIs are genuinely independent and there is no systemic bias, and in fact drops down for high $\rho$ and very low $q$. Extension of this work to cases where there is systematic bias is the subject of a paper in progress.

What can we say about P(q) for a single AI, which we'll denote by $P_1(q)$ to avoid confusion?

In all cases, if q= 100% or epsilon has zero variance then $P_1(q) = 1$, and $P_1(q)$ doesn't depend on c.

## Approximate behaviour of $P_1(q)$ with theoretical distributions

The value of $P_1(q)$ for intermediate values depends of course on the distribution v and ε and on how they are correlated. The simplest[1] measure of this is the Pearson Correlation Coefficient ρ, where:

$$\rho = \frac{a_i \sigma_v,}{\sqrt{(a_i^2 \sigma_v^2 + \sigma_\epsilon^2)}}$$

Broadly speaking it doesn't make very much difference what the distribution of v is for q>20%, but for q<20% the impact diverges sharply. For very fat-tailed distributions $P_1(q)$ increases with low q (because there are a few big outliers) whereas for a normal distribution $P_1(q)$ decreases q super-linearly (see Fig 1, and the Figures in the SI for other values of ρ)

Empirically $P_1(20\%)$ has much the same value for a given value of rho across each of these distributions. And a pretty good empirical approximation to $P_1(20\%)$ is:

$$P(20\%) \approx 0.2 + 0.5\rho + 0.3\rho^{10} \qquad (5)$$

Which has the correct values for ρ= 0 and ρ= 1. Thus for q>0.2 we can say:

$$P(q) \approx q + (1 - q)(0.625\rho + 0.375\rho^{10}) \qquad (6)$$

Or more crudely, which may be good enough for most purposes:

$$P(q) \approx q + (1 - q)(0.6\rho + 0.4\rho^{10}) \qquad (6a)$$

For q<20% the behaviour of the distributions diverges, but we can approximate to P(q) by a straight line in log space (*ie* a power law) towards P(1/m). The Normal Limit (red cross) is an exact calculation, the Student t-limit is based on a self-contained simulation, and the Heavy Tail estimate is a somewhat pragmatic approximation[2]. Since in reality neither the signal nor the noise will conform to exactly the assumed distributions, these approximations are probably sufficient. However it would be very unwise to assume that, in practice, $P_1(q)$ increases substantially as q reduces even in fat-tailed distributions, because this depends on the noise being Normally distributed and additive, which may not be the case.

[1] In real situations this single global measure may be too simplistic. It is possible for example that the correlation between the v & ε may vary due to some other property of the CVs.

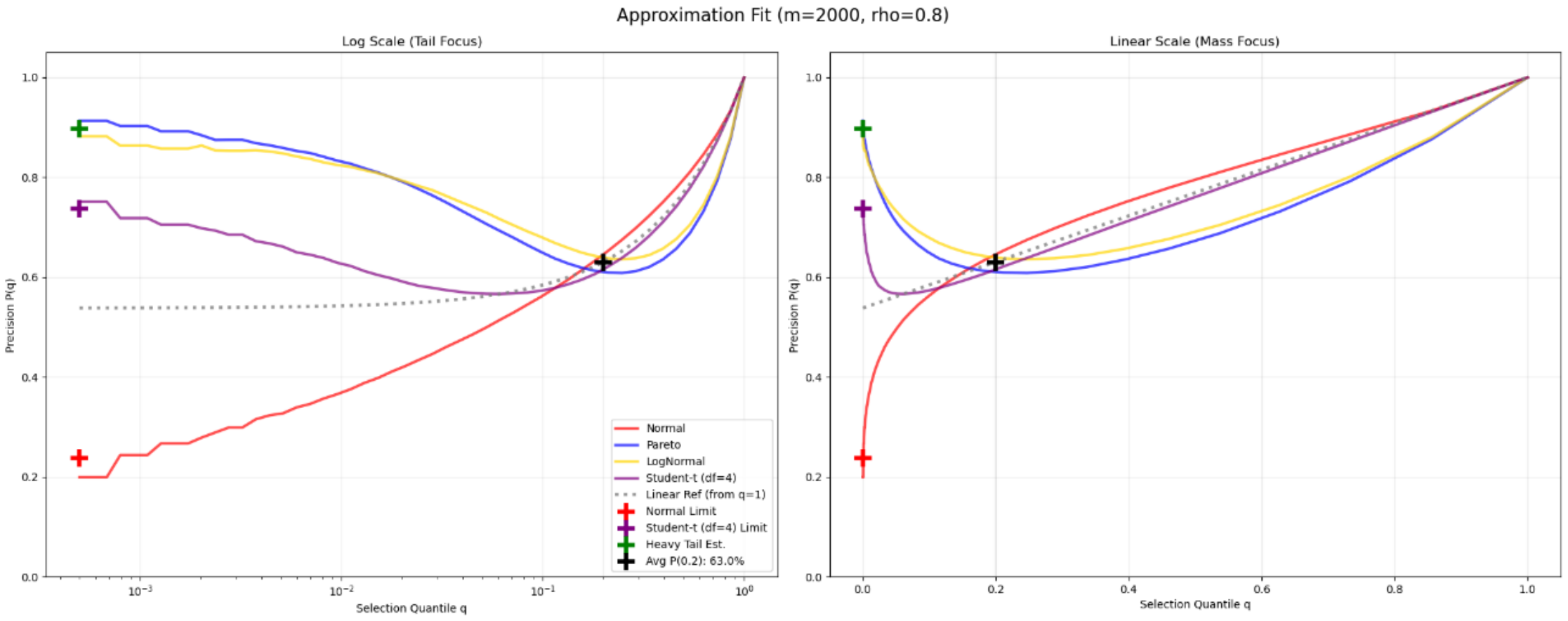

**Fig 1**: $P_1(q)$ as a function of rho for selected distributions. Note that $P_1(20\%)$ is pretty similar.

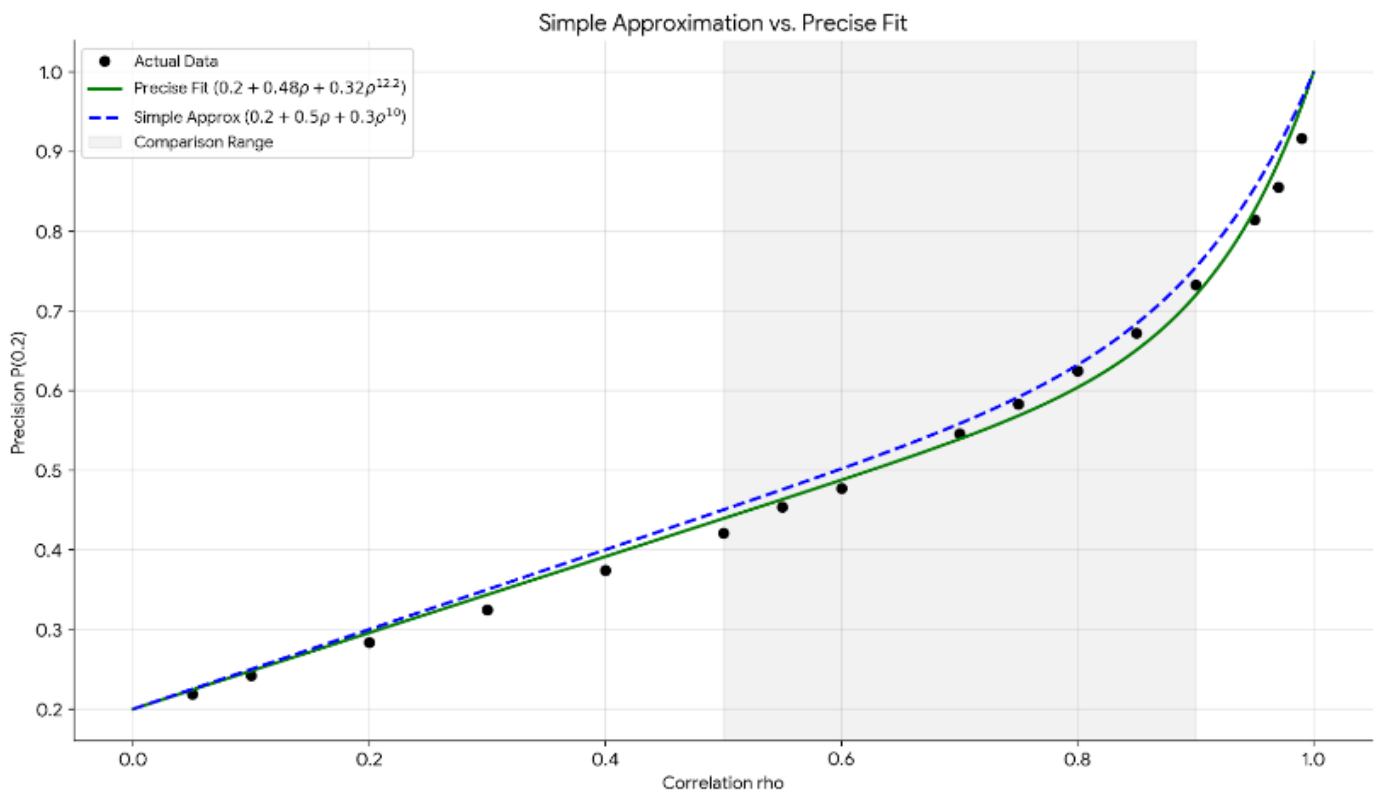

**Fig 2**: $P_1(20\%)$_bar as a function of rho. The theoretical best fit is close enough to the 1-sig fig version.

## Engaging with real data

To explore this in practice we collected a set of 200 outline CVs of Directors of UK-listed Public Companies based on information publicly announced by the London Stock Exchange Regulatory News Service (RNS). Because this was information required to be made public there were no privacy issues. We then asked 5 AIs to assess their suitability for a hypothetical Non-executive Director (NED) position on five of the largest UK listed companies: AstraZeneca, Shell, Rolls-Royce, HSBC and Unilever. These companies were chosen because they have a high public profile (so that the AIs would have a lot of information about them), they span a variety of sectors and are all amongst the most valuable companies in the UK. None of them was involved in the work. We asked the AIs to assess the suitability of each candidate for an NED role in each company on a scale of 0.0 to 10.0. We did not specify the type of NED role involved and in practice some NED have designated positions such as Chair of the Audit Committee and there is usually a specific brief attached to an NED recruitment. But most people who have served on a UK Listed Company board could fulfil and NED role on one of these boards *to some extent* and we wanted a plausible and rich set of data.

---

[2] The asymptotic value of P(0) in this case is 1, so we assume that P(1/10m) = 1 and linearly interpolate between this and the average P(0.2) point. It's really not bad.

40% of the candidates were female and we are keenly interested in exploring issues of systematic bias in the use of AI to assess candidates. A separate paper is in preparation to explore this further, and we ignore the issue here, not because we don't consider it important, but because it will be the subject of a separate paper.

The average score was 7.25 with a Standard Deviation of 0.57, a minimum of 4.8 and a maximum of 9.9. There was no significant difference between the average score of Female and Male candidates (7.35 v 7.19). The distribution of scores awarded was pretty close to Normal (see individual QQ plots in the SI).

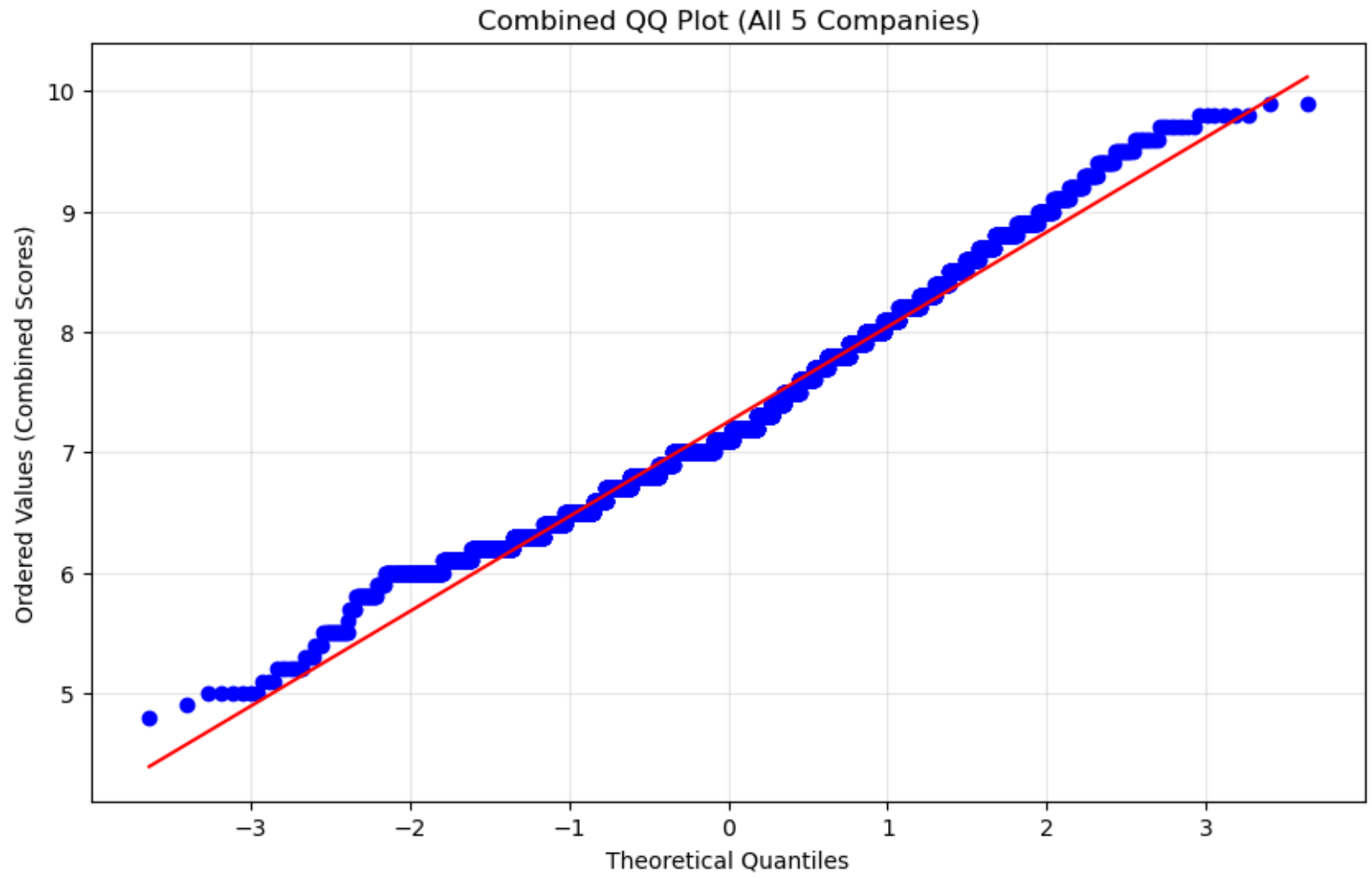

**Fig 3**: QQ Plot of AI Scores for all 5 Companies

To make a proxy *y* for the "true" score ν we find optimal weights $w_i$ summing to 1 so that $y_j = \Sigma w_i X_{ij}$

Using this as an approximate ground truth we can plot the estimated P(q) for each AI, and make an average (which is roughly the precision you could expect if you didn't know how good your particular AI was.

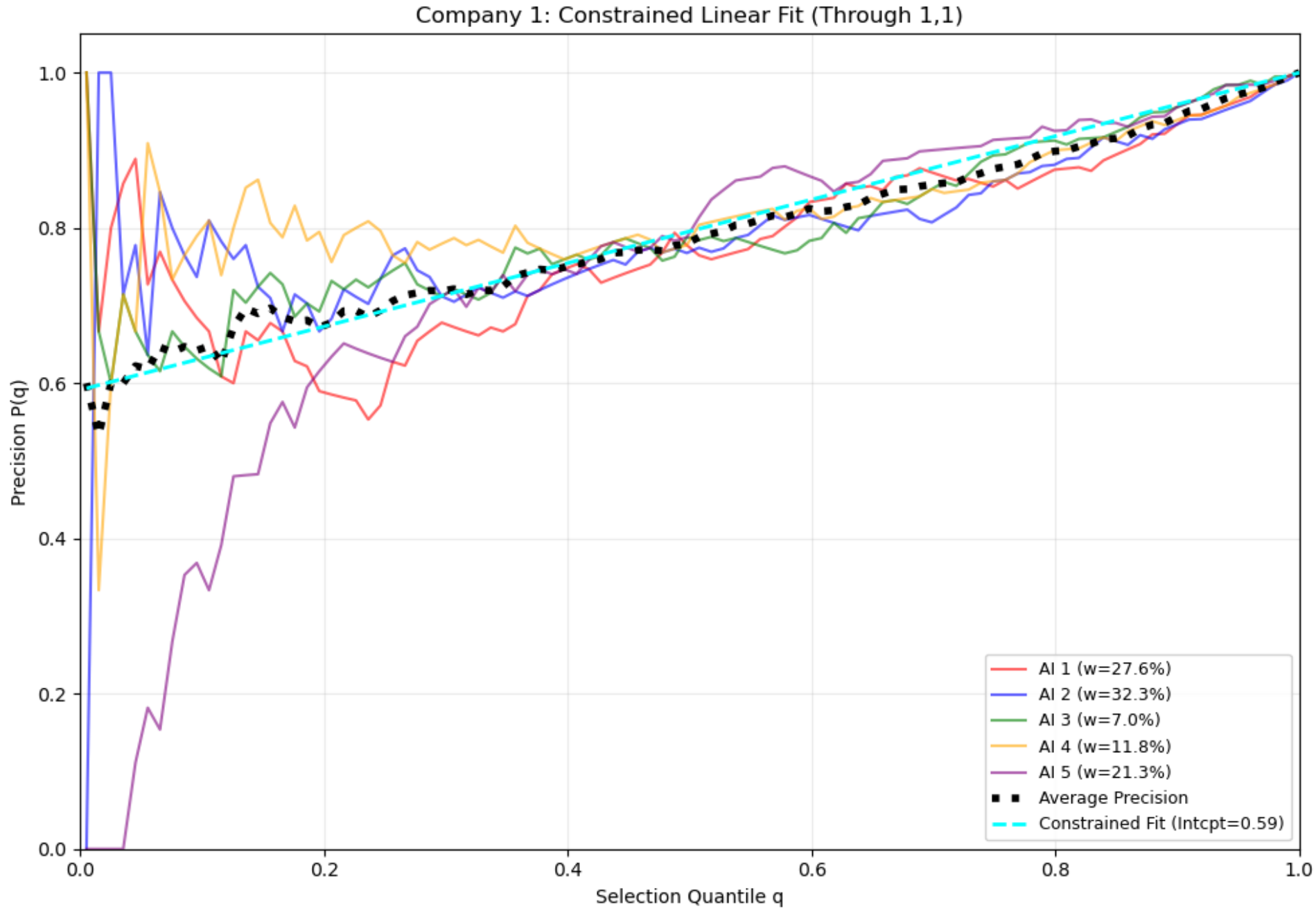

**Fig 4**: $P_1(q)$ for each AI for Company 1, average position & constrained fit.

What is quite striking is that the intercept of the light blue dotted line (which is the best fit to the average that goes through (1,1)) is roughly equal to the average correlation between the AIs. The results aren't perfect but are good enough to be usable to give a rough estimate:

$$P(q) \approx \rho + q\,(1 - \rho) \tag{2}$$

| Company | **1** | **2** | **3** | **4** | **5** |
|---|---|---|---|---|---|
| Average correlation | 0.545 | 0.618 | 0.493 | 0.471 | 0.539 |
| Intercept | 0.591 | 0.604 | 0.551 | 0.540 | 0.583 |
| % Difference | +8.5% | -2.2% | +10.6% | +14.7% | +8.3% |

This is closely related to a result of [Brogden 1946] where he showed that the Efficiency of a selection method was equal to its correlation coefficient, assuming that the predictor and ground truth are jointly normal. A simple derivation is to suppose that the AI is approximately a "Switching Mechanism" which is accurate with probability ρ and hallucinates randomly otherwise.

## Impact on Panel Size

We can now ask the following question: what is the impact of moving from 1 to 2 to 3 to 4 AIs in the "Panel"? This will give us a basis for some form of cost-benefit analysis. It will also help us understand how far our approximate ground truth is from a hypothetical ultimate ground truth *assuming* (as is almost certainly *not* the case) that there are no common systematic biases in all the AIs.

There are 10 possible pairs, 10 possible triples, and 5 possible quartets from the 5 AIs. The results of running them on the data for Company 1 are summarized in the graphs below – Companies 2, 3, 4 & 5 are shown in the SI.

One might expect the average precision of these panels to improve roughly accordance with the Spearman-Brown prediction formula [Brown 1910], published in 1910 in the context of psychometric testing. This is:

$$\mathrm{R}_n = \frac{n\bar{\rho}}{1+(n-1)\bar{\rho}} \tag{7}$$

Note the similarity to (4). There is reasonable empirical support for Equation (7) applying to our data (see 4th panel below). It also raises the question of just how far away our Approximate Ground Truth is.

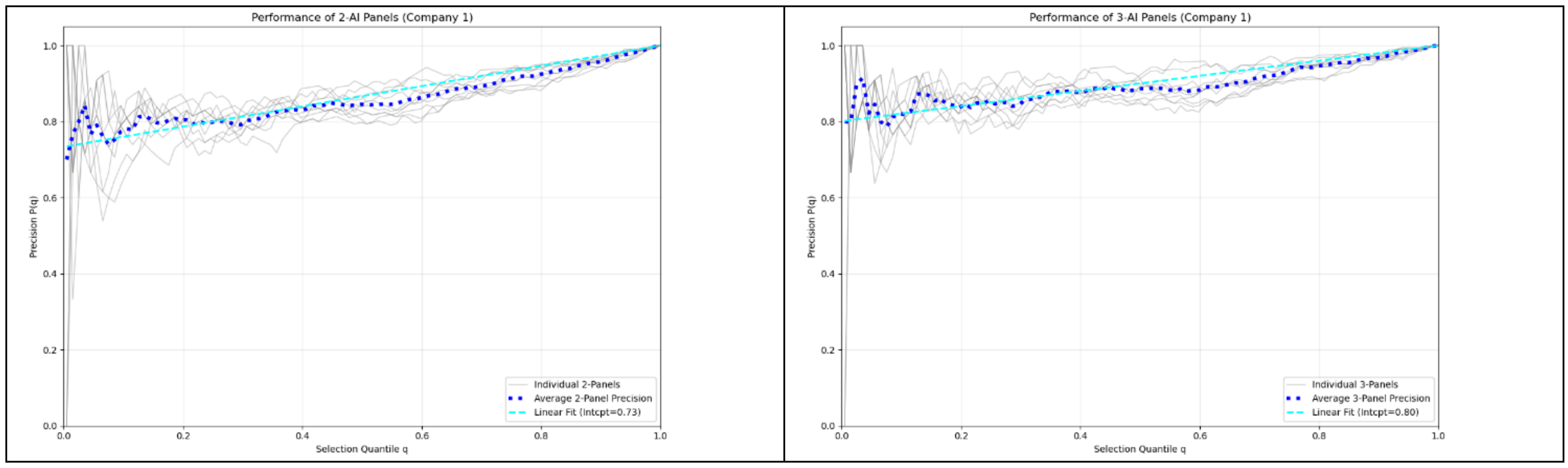

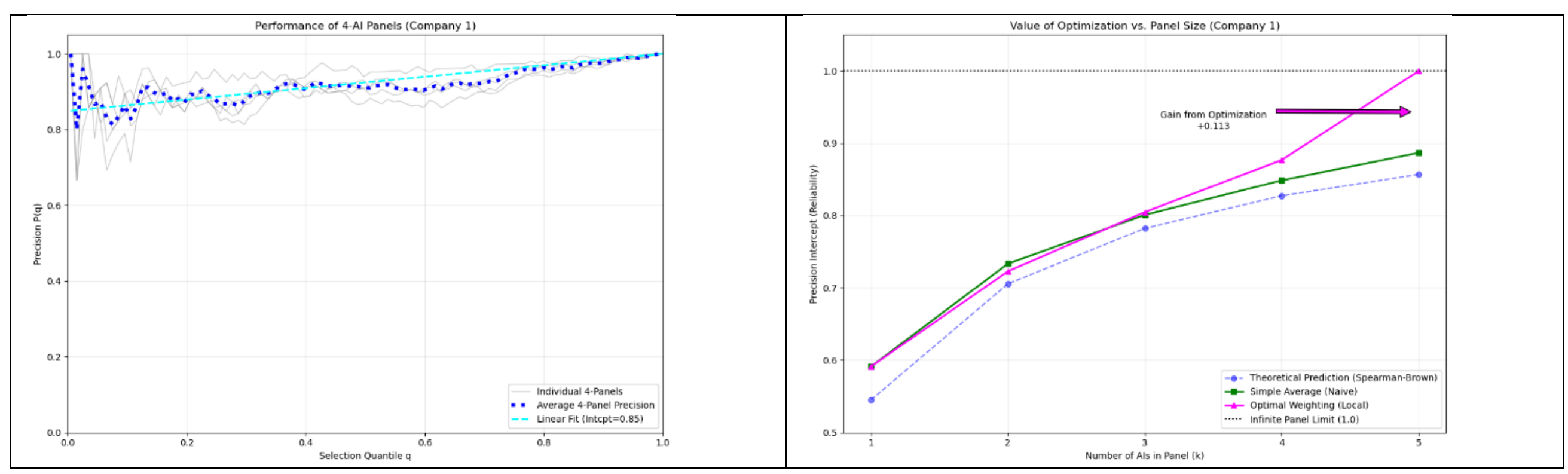

**Fig 5**: Use of Panel Sizes 2, 3 & 4 for Company 1 and Spearman-Brown predictions

| Panel Size | 1 | 2 | 3 | 4 |
|---|---|---|---|---|
| Observed  Avg Intercept | 0.591 | 0.734 | 0.801 | 0.849 |
| Spearman-Brown Prediction | | 0.706 | 0.783 | 0.828 |
| % Difference | | -3.6% | -2.3% | -2.5% |
| Observed % Improvement | | 24% | 9% | 6% |
| Predicted % Improvement | | 19% | 11% | 6% |

Comparable results for the other companies are given in the SI.

## Simulations of 'ideal' Panel

To dig deeper into the way in which additional AIs might improve the situation, we constructed simulated data with the same statistical characteristics as the observed data. To do this we had to use the Approximate Ground Truth distributions uncovered above. These are basically normal but have significantly fat tails at quantiles above +1.5 (see Fig 6 below).

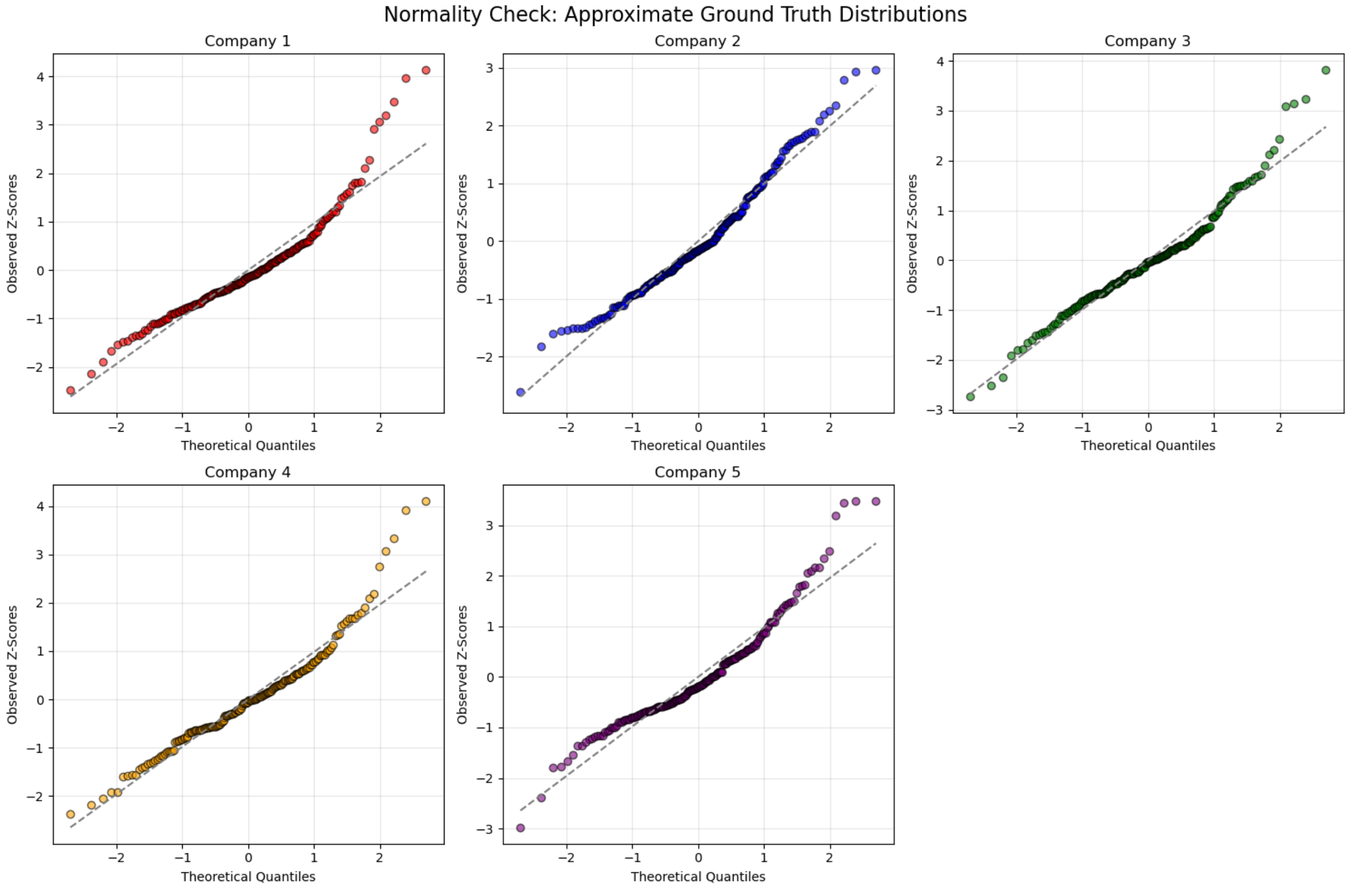

**Fig 6**: Quantiles of approximate "ground truth" estimated from the Datasets

We then synthesized a large number of datasets with similar characteristics (but with 2,000 candidates) and the "outputs" of 100 AIs that had random error injected in such a way as to achieve a distribution of correlations that mimicked the observed distributions of the correlations of the 5 AIs across 5 companies, which were pretty much Normally distributed (see SI). This allowed us to observe and quantify the scaling effect which we had observed, namely that the accuracy of Panel estimates did not grow as fast as [Brown 1910] predicted but as though there were an "effective number" of AIs which grew roughly as $n^{0.7}$. The qualitative explanation of this is that P(q) has a binary cutoff – a CV is either in the "true" quintile or it is not, and unlike the Reliability statistics in the pre-1950 literature[3] there is no credit for a 'near miss'. We went to considerable trouble to ensure that our samples were fat-tailed in a way that mimicked the data, but when we switched this effect off, the empirical values of *b* were essentially identical We are this confident that it is not an artefact to the details of the distributions. We are pretty sure that radically different distributions could be found where *b* behaved differently and would encourage others to look for them and for situations where they are appropriate. Figure 7 below shows the results of these simulations with 4000 runs.

| q q | Slope | Intercept | R-squared |
|---|---|---|---|
| 0.05 | -0.679 | 0.833 | 0.970 |
| 0.10 | -0.796 | 0.941 | 0.987 |
| 0.15 | -0.769 | 0.962 | 0.965 |
| 0.20 | -0.800 | 0.998 | 0.989 |
| 0.25 | -0.822 | 1.018 | 0.983 |
| 0.30 | -0.809 | 1.030 | 0.991 |

**Table 1: Calculated Intercepts**

Given that all assumptions about distributions are imprecise it seemed better to have a universal approximate formula, which is never very far out for reasonable *ρ* and q.

$$b \approx q + 0.8\,(1 - \rho) \qquad (8)$$

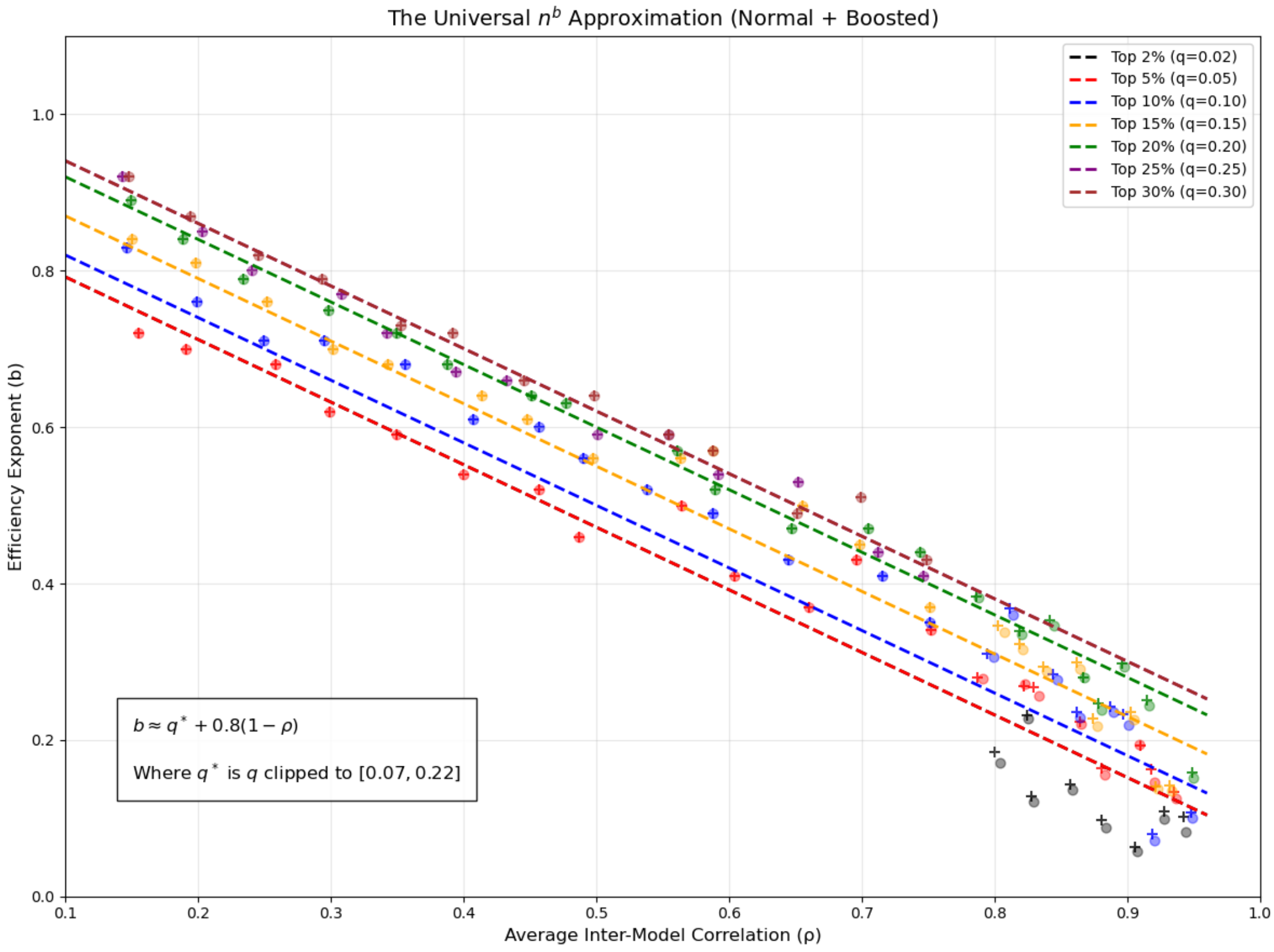

**Figure 7**: Observed Efficiency Exponents and approximate fit

[3] Of which of course we were wholly unaware until alerted by AI. Brogden 1946 has only been cited 12x since 2020.

Note that b drops off quite sharply when $\rho$ >0.9 but we know that the formula is an overestimate anyway and that when $\rho$ >0.9 the value added from larger panels is limited. We show for completeness the estimated values for b when q = 0.02. in the $\rho$ >0.8 zone. The RMS error is 0.043 and the maximum absolute error is 0.12. The key point is that this approximate formula should not be relied upon much when q is very small or $\rho$ very large. It is however a sensible approximation for reasonable values, when q is in the range 0.05-0.3. If a precise value of b is required it can be calculated by simulation once the distributions of the underlying population and noise terms are understood.

## Discussion

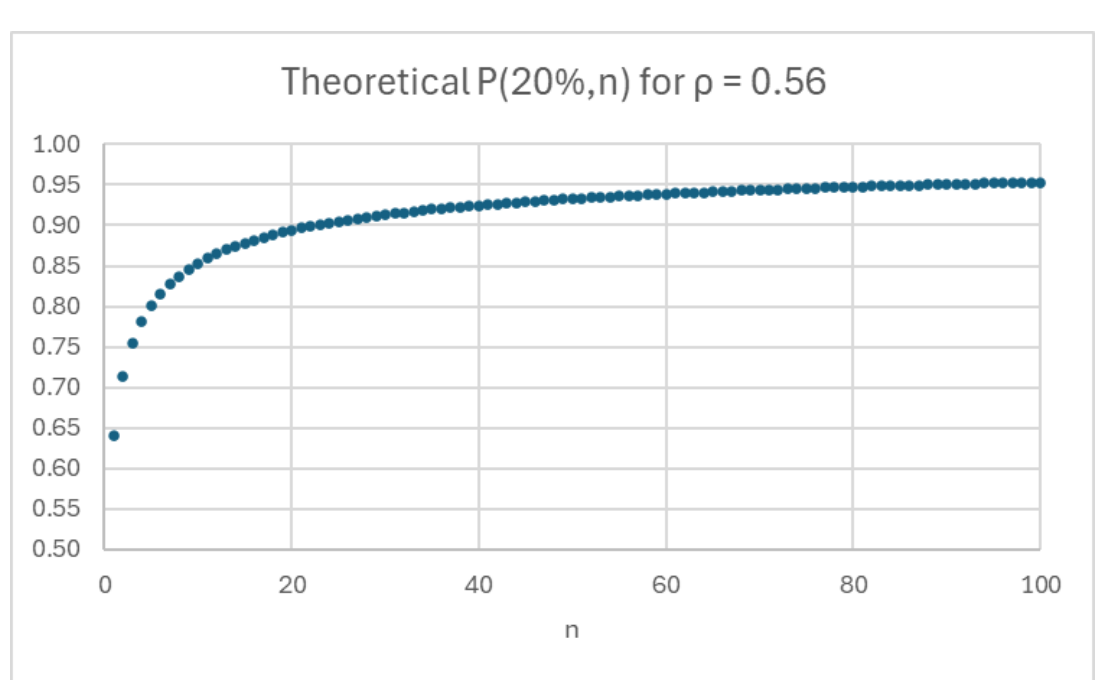

We have shown that for many practical applications of AI screening there is a clear advantage in using a panel of a reasonable number of AIs if you care about the result, and that it is possible to make reasonable estimates of the scope for improvement. For example in the datasets discussed, with an average correlation coefficient of c.0.55 P(20%) is about 64% for a single AI. This can be raised to c. 75% with a Panel of 3 AIs and c80% with a panel of 5 (b is c. 0.56) but getting very much above this is impracticable - even if a panel of 25 AI that didn't share systemic biases were available it would only achieve 90% with 85% at 10 AIs.

There are of course other important advantages in having diversity in the use of AIs. For example, "hallucinations" are generally considered to be largely random (in the sense that two different AIs will seldom "hallucinate" in the same way at the same time) and therefore it is easier to spot these with a panel of 3 or more AIs. In addition, if you only have 2-3 AIs you have very limited information about the true mean correlation between a hypothetical large panel and may therefore substantially over-estimate the average ρ and hence the reliability of your Panel's estimate. Generally speaking it would probably be wise to use a somewhat larger "Panel" than the calculations indicate, if the decision is at all important. And we would reiterate that increasing the Panel size does not in itself address issues of systematic bias across all AIs, although the separate paper in development addresses this issue to some extent.

It is also worth emphasizing that decisions about hiring are some of the most important strategic decisions that organizations ever make. This is arguably especially true of hiring for entry-level positions because an organization's longest serving staff typically begin as entry level hires (see *eg* [Becker 1962] but also [Vegetius 390], [Hall 1831] and many others). For organisations to rely on a single AI or, collectively, on a small set of such AIs risks reducing the diversity of such organisations and thus increasing systemic risks [ Beale & al, 2011]. There is also the serious additional problem that the AIs offered to organisations for this purpose may in some cases be attempting to maximise profits for their owners or developers and thus susceptible to issues around Unethical Optimization [Beale & al 2020].

## Conclusion

In this paper we have shown quantitatively that AI Panels improve precision and given a practical formula for estimating the precision gain. We see this both as a specific practical contribution and a corrective against an ideology/mythology which sees AIs as horses on a race towards some kind of omniscience, which for reasons discussed in scientifically unachievable. That computers can perform *some* intellectual tasks better than any human being is not doubt. But outside some quite narrow domains such as mathematics, great intellectual abilities do not necessarily result in agreement or wisdom. If AIs do become "god-like" they are likely to resemble the Homeric gods, and in the classical world the wise citizen hedged their bets and sought the protection of more than one of the often quarrelling deities [see *eg* Gibbon 1789, Beard 2015].

## Future Work

Other potentially fruitful areas for further work would include explicitly modelling the uncertainties around assessments, especially since AIs are probabilistic and can be asked to disclose their levels of uncertainty, which could include building on the work of [MacKay & al, 2017]. We have in fact done some work of this type, asking AIs to be explicit about their confidence intervals, recalibrating their over-confidence, and then estimating a combined distribution. It is also straightforward to get some measure of the divergence between a Panel of AIs on eg a CV by observing the standard deviation of the estimates. It would also be interesting to extend the analysis to address the fact that the real purpose of most AI-driven pre-screening is to get a manageable "long list" from which a more careful selection can be made. One could define:

$$P(h, q; x, \nu) \coloneqq \frac{Card\big(Top(h, \nu) \cap Top(q, x)\big)}{hm}$$

So P(10%,25%) would be the probability that someone who was really in the top 10% (of ν) would be selected as being in the top 25% of x. This could be a fruitful line of further enquiry.

However the intriguing mathematics should not blind us to wider systemic and socioeconomic issues. One of these is that the use of a relatively small number of AI systems to filter job candidates greatly reduces the diversity of the filtering. The impact of some candidates being unfairly excluded from consideration by one employer is mitigated if there are a large number of alternative employers who will not exclude these candidates. But if all employers use similar filtering criteria they may unfairly exclude similar sets of candidates – reminiscent of the "diverse diversification" issues in [Beale & al 2011].

In addition there will be contexts where it may be better to use a median or a clipped mean rather than an arithmetic mean, so that a single AI which is "hallucinating" or for some reason has a specific bias for or against an individual is disregarded. We also need to consider how to address AIs which have the equivalent of "sponsored items" where companies pay to influence the recommendations that the AI makes [Rijo, 2026]. Even without that, some companies are already giving careful thought to how they can optimize their internet profile to maximize the chances of AIs responding favourably to questions about them [Aggarwal & al 2024], and Adversarial Optimization can be applied to CVs to raise the scores that AIs give to them either with blatantly nefarious approaches [Mu & al 2025] [Akdemir & Levy 2025] or by more traditional approaches of injecting "non-random noise" such as TextFooler [Jin & al 2020]. It is reasonable to conjecture that a genuinely diverse panel of AIs would be more robust against such techniques. Attacks against ensembles of AI classifiers have been studied for many years [eg Tramèr & al 2017] and it would certainly be interesting to develop an "adversarial diversity measure" which looked at the differential susceptibility of different AI models to specific adversarial optimisations, to complement the correlation-based diversity measures used in this paper.

To conclude: we believe that a quantitative discussion of the merits of using a diverse panel of AIs to support decision-making in such areas as CV screening will move people away from dangerous reliance on single AI systems and encourage a balanced assessment of the extent to which diversity needs to be built into the AI parts of the socioeconomic systems that are so important for our future.

## References

- [Aggarwal & al 2024] **Aggarwal, P., Murahari, V., Rajpurohit, T., Kalyan, A., Narasimhan, K., and Deshpande, A**. 2024. GEO: Generative Engine Optimization. In *Proceedings of the 30th ACM SIGKDD*

*Conference on Knowledge Discovery and Data Mining* (KDD '24). Association for Computing Machinery, New York, NY, USA, 5–16. https://doi.org/10.1145/3637528.3671900

- [Akdemir & Levy 2025] Akdemir, A. and Levy, J. Understanding and Defending Against Resume-Based Prompt Injections in HR AI *RecSys in HR'25: The 5th Workshop on Recommender Systems for Human Resources,* in conjunction with the 19th ACM Conference on Recommender Systems, September 22–26, 2025, Prague, Czech Republic.
- [Avižienis, 1967] **Avižienis, A**. Design of fault-tolerant computers *AFIPS '67 (Fall): Proceedings of the November 14-16, 1967, fall joint computer conference* pp733 – 743 doi.org/10.1145/1465611.14657
- [Beale & al 2011] **Beale, N., Rand, D. G., Battey, H., Croxson, K., May, R. M. & Nowak, M. A.** Individual versus systemic risk and the Regulator's Dilemma, *Proc. Natl. Acad. Sci. U.S.A.* 108 (31) 12647-12652, https://doi.org/10.1073/pnas.1105882108 (2011).
- [Beale & al 2020] **Beale, N. Battey, H. Davison, A. C., MacKay, R. S**.; An unethical optimization principle. *R Soc Open Sci.* 1 July 2020; 7 (7): 200462. https://doi.org/10.1098/rsos.200462
- [Beard, 2015] **Beard, M.** *SPQR: A History of Ancient Rome* London. Profile Books ISBN 978-0-87140-423-7
- [Becker 1962] **Becker, G. S.** (1962). "Investment in Human Capital: A Theoretical Analysis." *Journal of Political Economy*.
- [Brown 1910] **Brown, W.** Some experimental results in the correlation of mental abilities. *British Journal of Psychology, 3*, 296–322. (1910)
- [Brogden 1946] **Brogden, H. E.** "On the interpretation of the correlation coefficient as a measure of predictive efficiency." *Journal of Educational Psychology, 37*(2), 65–76. (1946)
- [Caliskan & al 2017] **Caliskan, A., Bryson, J. J., & Narayanan, A**. (2017). *Semantics derived automatically from language corpora contain human-like biases*. *Science*, 356(6334), 183–186
- [Chen, 2023] **Chen, Z**. "Ethics and discrimination in artificial intelligence enabled recruitment practices" *Humanities and Social Sciences Communications* (2023) 10:567 doi.org/10.1057/s41599-023-02079-x.
- [Ferrara 2024] **Ferrara, E**. Fairness and Bias in Artificial Intelligence: A Brief Survey of Sources, Impacts, and Mitigation Strategies. *Sci*. 2024; 6(1):3. https://doi.org/10.3390/sci6010003.
- [Gibbon 1776-89] **Gibbon, E.** *The History of the Decline and Fall of the Roman Empire* Originally published in London by Strahan & Cadell 1776-1789. One modern edition is ISBN 0-679-42308-7 (vols. 1–3); ISBN 0-679-43593-X (vols. 4–6). Although his details and interpretations are questionable he wrote matchless prose. For example "The various modes of worship, which prevailed in the Roman world, were all considered by the people, as equally true; by the philosopher, as equally false; and by the magistrate, as equally useful." (Ch 2)
- [Gödel, 1931] **Gödel, K.** Über formal unentscheidbare Sätze der Principia Mathematica und verwandter Systeme I *Monatshefte für Mathematik und Physik* 1931
- [Hall 1831] **Hall, B.** *Fragments of Voyages and Travels*, 1831. "It is in the midshipmen's berth that the officers of the navy are formed."
- [Hong & Page, 2004] **Hong, L & Page S.E.** Groups of diverse problem solvers can outperform groups of high-ability problem solvers, *Proc. Natl. Acad. Sci. U.S.A.* 101 (46) 16385-16389, (2004) https://doi.org/10.1073/pnas.0403723101.

- [Jin & al 2020] **Jin, D., Jin,Z., Zhou,J. Szolovits, P.,** "Is BERT Really Robust? A Strong Baseline for Natural Language Attack on Text Classification and Entailment," *AAAI 2020*, arXiv:1907.11932.
- [Lakshminarayanan et al. 2017] **Lakshminarayanan, B. Pritzel, A. Blundell, C.** Simple and Scalable Predictive Uncertainty Estimation using Deep Ensembles NeurIPS 2017, arXiv:1612.01474
- [Li & al 2025] **Li, W. Lin, Y. Xia, M. Jin C.** Rethinking Mixture-of-Agents: Is Mixing Different Large Language Models Beneficial? arXiv:2502.00674
- [Lucas 1961] **Lucas, J. R.** Minds Machines and Gödel *Philosophy XXXVI* 1961. This is the printed copy of the paper he read in 1959 to the Oxford Philosophical Society. See also [Lucas 1970]
- [Lucas 1970] **Lucas, J. R.** *The Freedom of the Will* Oxford University Press 1970 ISBN 978-0198243434
- [Mallaby, 2026] **Mallaby, S.** *The Infinity Machine: Demis Hassabis, Deepmind and the Quest for Superintelligence* Penguin Random House 2026. “A Council on Foreign Relations Book”. Mallaby does mention Gödel because of course Hassabis is deeply aware of him, but doesn’t appreciate the implications, speaking of neural networks processing “a near infinity of bits…disproving…claims about the limits of classical computers”. (p392)
- [MacKay & al 2017] **MacKay, R. S., Kenna, R., Low, R. J., Parker, S.**; Calibration with confidence: a principled method for panel assessment. *R Soc Open Sci.* 1 February 2017; 4 (2): 160760. https://doi.org/10.1098/rsos.160760
- [Mu & al 2025] **Mu, H., Liu, J., Wan, K., Xing, R., Chen, X., Baldwin, T., Che, W.** AI Security Beyond Core Domains: Resume Screening as a Case Study of Adversarial Vulnerabilities in Specialized LLM Applications arXiv:2512.20164v1 [cs.CL] 23 Dec 2025
- **[**Penrose 1989] **Penrose, R.** *The Emperor’s New Mind* Oxford, OUP ISBN 0-19-851973-7 builds on Lucas’ argument and adds some intriguing speculation about quantum mechanical collapse, which is not relevant to the present paper.
- [Raghavan & al 2020] **Raghavan, M., Barocas, S., Kleinberg, J., and Levy, K**. 2020. Mitigating Bias in Algorithmic Hiring: Evaluating Claims and Practices. In Conference on Fairness, Accountability, and Transparency (FAT* ’20), January 27–30, 2020, Barcelona, Spain. ACM, New York, NY, USA, 13 pages. doi.org/10.1145/3351095.3372828
- [Rijo 2026] **Rijo, L.** Sponsored stores and quick web results spotted inside Google AI Mode. *PPC Land* 6 April 2026. https://ppc.land/sponsored-stores-and-quick-web-results-spotted-inside-google-ai-mode/ See also https://www.airanklab.com/blog/ai-search-state-of-market-report for discussion of this trend in other AIs.
- [Schapire, 1990] **Schapire, R. E**. The Strength of Weak Learnability," *Machine Learning* 5(2):197–227 doi.org/10.1007/BF00116037
- [Tramèr & al 2017] **Tramèr, F., Kurakin, A., Papernot, N., Goodfellow, I., Boneh, D., & McDaniel, P.** Ensemble Adversarial Training: Attacks and Defenses. This went up on arXiv in 2017, was published in *ICLR 2018* and revised in 2020 arXiv:1705.07204v5 [stat.ML] 26 Apr 2020
- [Vegetius 390] **Vegetius, P. F. R.**, *De Re Militari*, Book I, “the safety of the whole Republic depends on the choice of recruits.” (c. 390).
- [Wang & al 2025] **Wang, J. Wang, J., Athiwaratkun, A. Zhang, C., Zhou, J.** Mixture-of-Agents Enhances Large Language Model Capabilities arXiv:2406.04692v1 [cs.CL] 7 Jun 2024

## Miscellaneous

**Use of AI**: As indicated, five AI systems were used to generate the data in early 2025, based on RNS disclosures. The entire dataset is on an excel spreadsheet which can be deposited.

I used Google Gemini 3 as an "unreliable research assistant" to generate the python code and as a virtual discussion partner in the mathematics. There were several cases in which I had to correct the code Gemini produced. I then ran the code on my own Python/Jupyter environment. AI was not used to write the paper. Gemini 3 suggested keywords which I then edited, and suggested references – at least one of which turned out to be a hallucination. Preflight was used to suggest format and editorial improvements. A couple of references were found by ChatGPT but I checked them and read the papers. Following an initial rejection on the grounds that the paper lacked sufficient Introduction, Background, Discussion, Conclusion and Future Work sections I used Claude Sonnet 4.6 as a 'discussion partner' to suggest areas of improvement, help identify additional references, and review the improved texts. I naturally checked all the references and read the papers (which was absolutely necessary – AIs are by no means entirely reliable) and wrote all the additional text.

**Author Contribution**: I wrote the paper and ran the software to perform the calculations.

**Acknowledgements**: I have benefited from many helpful discussions, including with Howard Haughton (KCL), Martine Barons (Warwick), Aniko Eckart (Aston), Robert Mackay FRS (Warwick), Heather Battey (Imperial), Anthony Davison (EPFL) and Wendy Hall FRS (Southampton).

**Conflicts of interest**: I identify no conflicts of interest. Naturally we use Panels of AI in our work at Sciteb.

**Funding**: I have received no external funding for this work.

**Ethics**: As noted above, because the personal data used has been publicly disclosed in accordance with legal requirements by the relevant companies, there are no ethical issues in using the data.

**Data access**: All the data and code will be available in an appropriate repository.

# Supplementary Information

## Charts for other values of rho and m

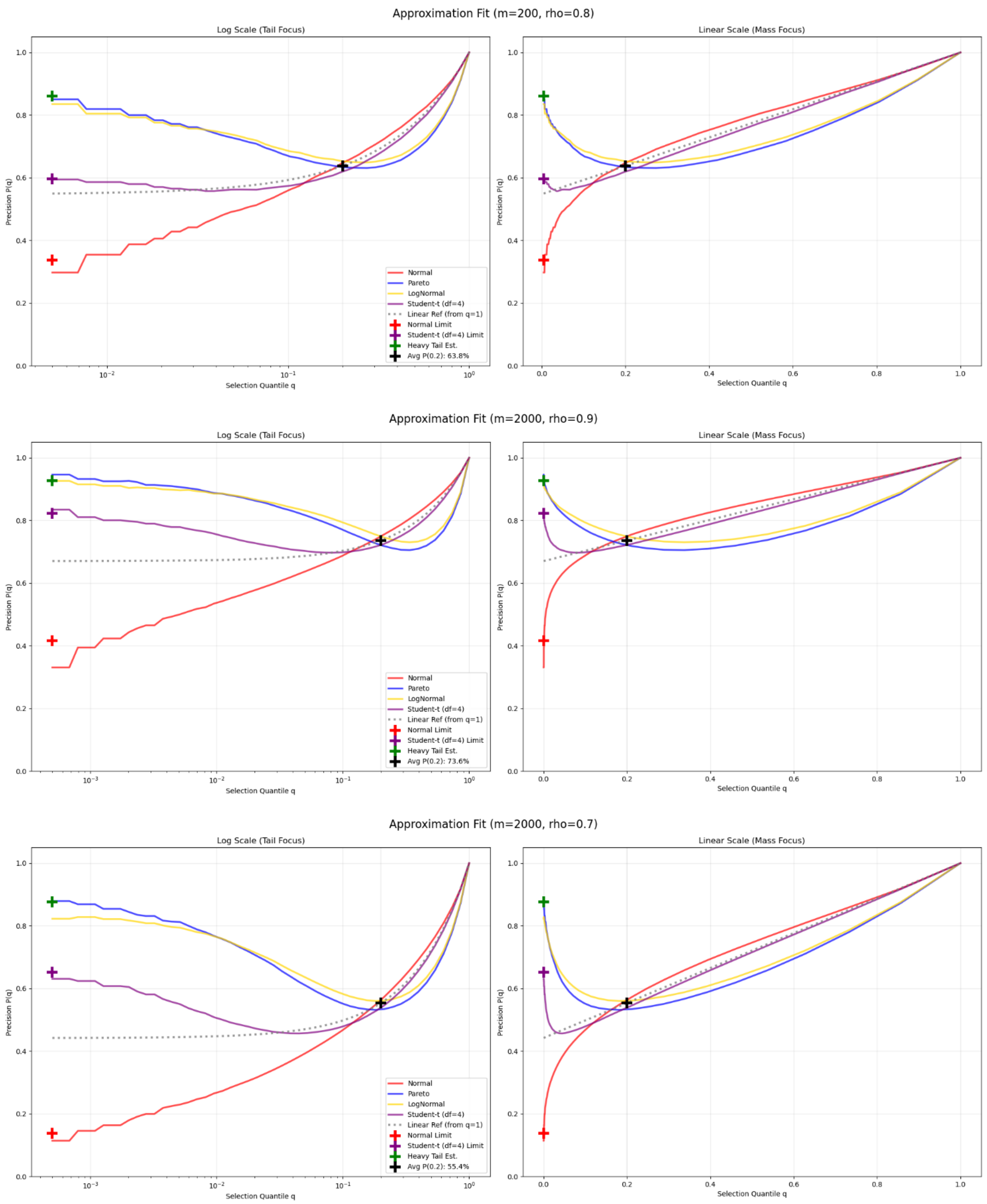

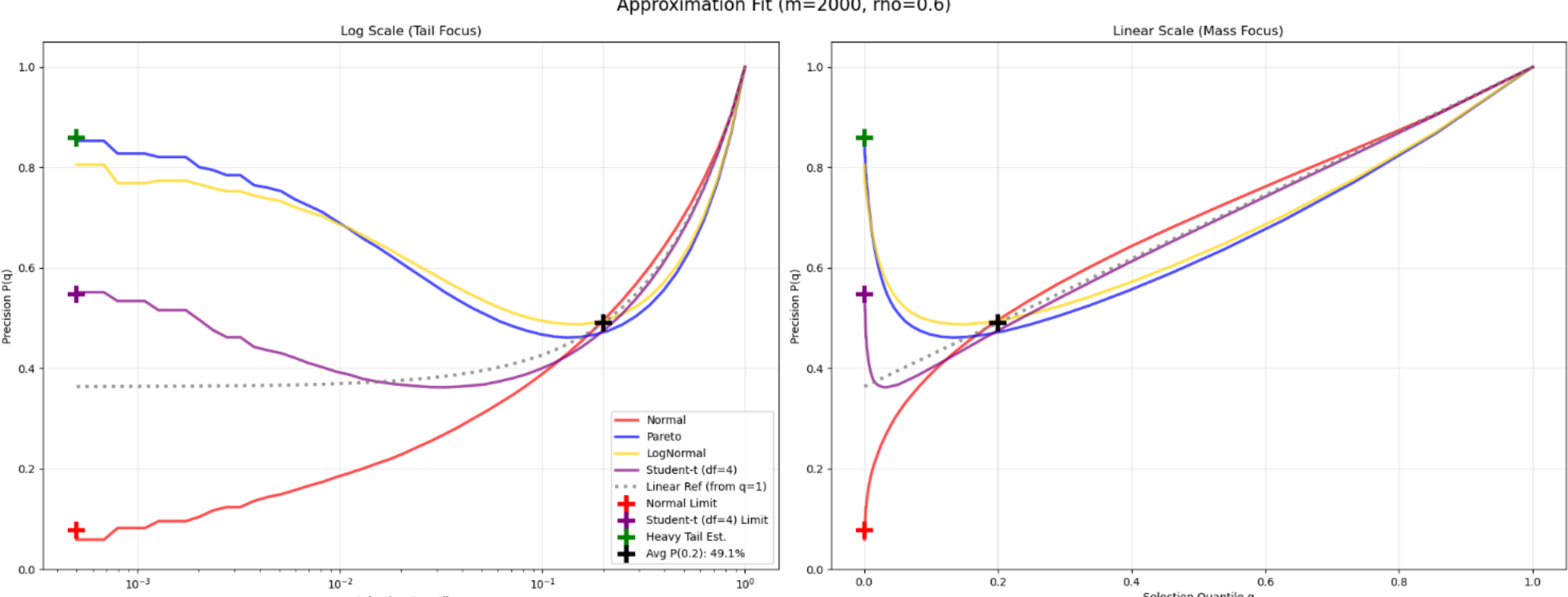

## Code for these calculations

```
import numpy as np
import scipy.stats as stats
import pandas as pd
import matplotlib.pyplot as plt
from scipy.stats import multivariate_normal, norm

# ==========================================
# CONFIGURATION
# ==========================================
VERSION = "Lansdowne Visualizer v2.07 (Sample-Aware Anchors) 30 Dec 2025"

# ==========================================
# 1. CORE FUNCTIONS
# ==========================================
def generate_scores(n, dist_type, rho_target, t_dof=4):
    """
    Generates scores for a specific distribution.
    Uses Sample Standard Deviation to scale noise.
    """
    if dist_type == 'Normal':
        nu = np.random.normal(0, 1, n)
    elif dist_type == 'LogNormal':
        nu = np.random.lognormal(0, 1, n)
    elif dist_type == 'Pareto':
        nu = stats.pareto.rvs(3, size=n)
    elif dist_type == 'Student-t':
        nu = stats.t.rvs(t_dof, size=n)
    else:
        raise ValueError(f"Unknown dist: {dist_type}")

    # Additive Normal Noise scaled by SAMPLE standard deviation
    sigma_nu = np.std(nu)
    noise_scale = sigma_nu * np.sqrt((1/rho_target**2) - 1)
    epsilon = np.random.normal(0, noise_scale, n)
    return nu, nu + epsilon

def calculate_t_anchor_finite(m, rho, dof, n_trials=50_000):
    """
    Calculates P(1/m) for the 'Signal + Noise' model by strictly simulating
    finite-sample batches. This corrects the 'Population vs Sample' variance bias.
    """
    # We use a vectorized approach to simulate n_trials at once for speed
    # Shape: (n_trials, m)

    # 1. Generate Signal Matrix
    nu = stats.t.rvs(dof, size=(n_trials, m))

    # 2. Calculate Sample Stds per trial (axis 1)
    sigmas = np.std(nu, axis=1, keepdims=True)

    # 3. Generate Noise scaled by specific trial sigma
```

```
    noise_scales = sigmas * np.sqrt((1/rho**2) - 1)
    epsilon = np.random.normal(0, 1, size=(n_trials, m)) * noise_scales

    # 4. Observed Scores
    x = nu + epsilon

    # 5. Find Top 1 Indices
    # argmax gives the index of the highest observed score in each trial
    winner_indices = np.argmax(x, axis=1)

    # 6. Check if Winner is True Top 1
    # We need to check if the winner's true talent (nu) is the max in that row
    # Fancy indexing to get the true talent of the winners
    winner_true_talents = nu[np.arange(n_trials), winner_indices]

    # Find the max true talent in each row
    max_true_talents = np.max(nu, axis=1)

    # Precision = fraction of trials where Winner Talent == Max Talent
    # (Using a small tolerance for float equality, though unlikely to be issue with continuous vars)
    hits = np.sum(winner_true_talents >= max_true_talents)

    return hits / n_trials

def calculate_anchors(m, rho, p_avg_02, t_dof):
    q = 1/m

    # --- Red: Normal (Exact CDF) ---
    z = norm.ppf(1 - q)
    cdf_val = multivariate_normal.cdf([z, z], mean=[0,0], cov=[[1, rho], [rho, 1]])
    p_norm = (1 - 2*(1-q) + cdf_val) / q

    # --- Purple: Student-t (Finite Sample Corrected) ---
    p_t = calculate_t_anchor_finite(m, rho, t_dof)

    # --- Green: Heavy Tail Est ---
    log_start = np.log10(1/(10*m))
    log_end   = np.log10(0.2)
    log_target= np.log10(1/m)
    fraction = (log_target - log_start) / (log_end - log_start)
    p_heavy_est = 1.0 - fraction * (1.0 - p_avg_02)

    return p_norm, p_t, p_heavy_est

# ==========================================
# 2. MAIN SIMULATION ENGINE
# ==========================================
def run_lansdowne_simulation(m=2000, rho=0.8, trials=2000, t_dof=4):

    print(f"{VERSION}")
    print(f"Parameters: m={m}, rho={rho}, t_DoF={t_dof}, trials={trials}")
    print("Running simulation... (Approx 15-20s)")
    print("-" * 80)

    distributions = ['Normal', 'Pareto', 'LogNormal', 'Student-t']
    colors = {
        'Normal': 'red',
        'Pareto': 'blue',
        'LogNormal': 'gold',
        'Student-t': 'purple'
    }

    q_calc = np.logspace(np.log10(1/m), 0, 50)
    plot_data = {dist: np.zeros(len(q_calc)) for dist in distributions}

    # --- A. Simulation Loop ---
    for t in range(trials):
        for dist in distributions:
            nu, x = generate_scores(m, dist, rho, t_dof)

            sort_idxs = np.argsort(-x)
            true_ranks = stats.rankdata(-nu, method='ordinal')

            for i, q in enumerate(q_calc):
                k = max(int(np.round(q * m)), 1)
                picked_indices = sort_idxs[:k]
                hits = np.sum(true_ranks[picked_indices] <= k)
```

```
                plot_data[dist][i] += (hits / k)

    for dist in distributions:
        plot_data[dist] /= trials

    # --- B. Reference Lines ---
    idx_02 = np.abs(q_calc - 0.2).argmin()
    p02_values = [plot_data[dist][idx_02] for dist in distributions]
    p_avg_02 = np.mean(p02_values)

    slope = (1.0 - p_avg_02) / 0.8
    grey_line_y = 1.0 + slope * (q_calc - 1.0)

    # --- C. Calculate Anchors ---
    p_norm_anc, p_t_anc, p_heavy_anc = calculate_anchors(m, rho, p_avg_02, t_dof)

    # --- D. Print Summary ---
    print(f"avg P(0.2) = {p_avg_02:.1%}, Normal Limit = {p_norm_anc:.1%}, "
          f"Student t (df={t_dof}) Limit = {p_t_anc:.1%}, Heavy Tailed Limit = {p_heavy_anc:.1%}")

    # --- E. Plotting ---
    fig, (ax1, ax2) = plt.subplots(1, 2, figsize=(20, 8))
    axes = [ax1, ax2]

    for i, ax in enumerate(axes):
        is_log = (i == 0)

        # 1. Curves
        for dist in distributions:
            lbl = dist
            if dist == 'Student-t':
                lbl = f"Student-t (df={t_dof})"
            ax.plot(q_calc, plot_data[dist], color=colors[dist], alpha=0.7, linewidth=2.5,
                    label=lbl if is_log else "")

        # 2. Grey Line
        ax.plot(q_calc, grey_line_y, color='grey', linestyle=':', linewidth=3, alpha=0.8,
                label="Linear Ref (from q=1)" if is_log else "")

        # 3. Anchors
        q_anchor = 1/m
        mk_size = 250
        mk_wid = 4

        ax.scatter([q_anchor], [p_norm_anc], color='red', marker='+', s=mk_size, linewidth=mk_wid,
                   zorder=10, label='Normal Limit' if is_log else "")
        ax.scatter([q_anchor], [p_t_anc], color='purple', marker='+', s=mk_size, linewidth=mk_wid,
                   zorder=10, label=f'Student-t (df={t_dof}) Limit' if is_log else "")
        ax.scatter([q_anchor], [p_heavy_anc], color='green', marker='+', s=mk_size, linewidth=mk_wid,
                   zorder=10, label='Heavy Tail Est.' if is_log else "")
        ax.scatter([0.2], [p_avg_02], color='black', marker='+', s=mk_size, linewidth=mk_wid,
                   zorder=10, label=f'Avg P(0.2): {p_avg_02:.1%}' if is_log else "")

        # Formatting
        ax.set_xlabel('Selection Quantile q')
        ax.set_ylabel('Precision P(q)')
        ax.set_ylim(0, 1.05)
        ax.grid(True, alpha=0.25)
        ax.axvline(x=0.2, color='gray', linestyle='-', alpha=0.1)

    ax1.set_xscale('log'); ax1.set_title("Log Scale (Tail Focus)")
    ax2.set_title("Linear Scale (Mass Focus)")
    ax1.legend(loc='lower right', framealpha=0.9, fontsize=10)

    plt.suptitle(f'Approximation Fit (m={m}, rho={rho})', fontsize=16)
    plt.tight_layout()
    plt.show()

# Run Production
run_lansdowne_simulation(m=2000, rho=0.8, trials=2000, t_dof=4)
```

## More

## QQ Plots for individual Companies

Combined QQ Plot (All 5 Companies)
Ordered Values (Combined Scores)
Theoretical Quantiles

QQ Plot of Actual AI Scores X1
Ordered Values
Theoretical quantiles

QQ Plot of Actual AI Scores X2
Ordered Values
Theoretical quantiles

QQ Plot of Actual AI Scores X3
Ordered Values
Theoretical quantiles

QQ Plot of Actual AI Scores X4
Ordered Values
Theoretical quantiles

QQ Plot of Actual AI Scores X5
Ordered Values
Theoretical quantiles

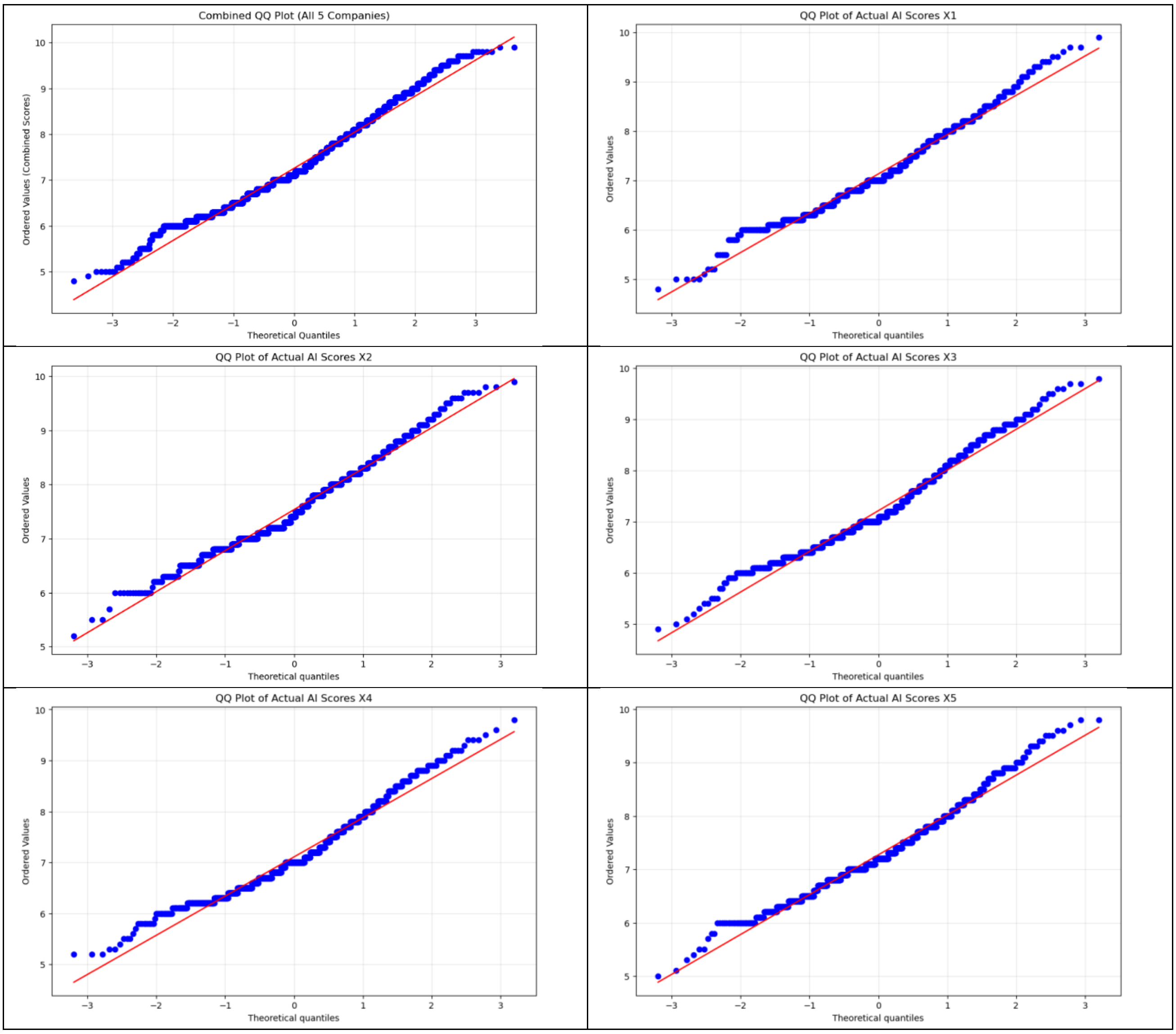

## Precision graphs for the other companies

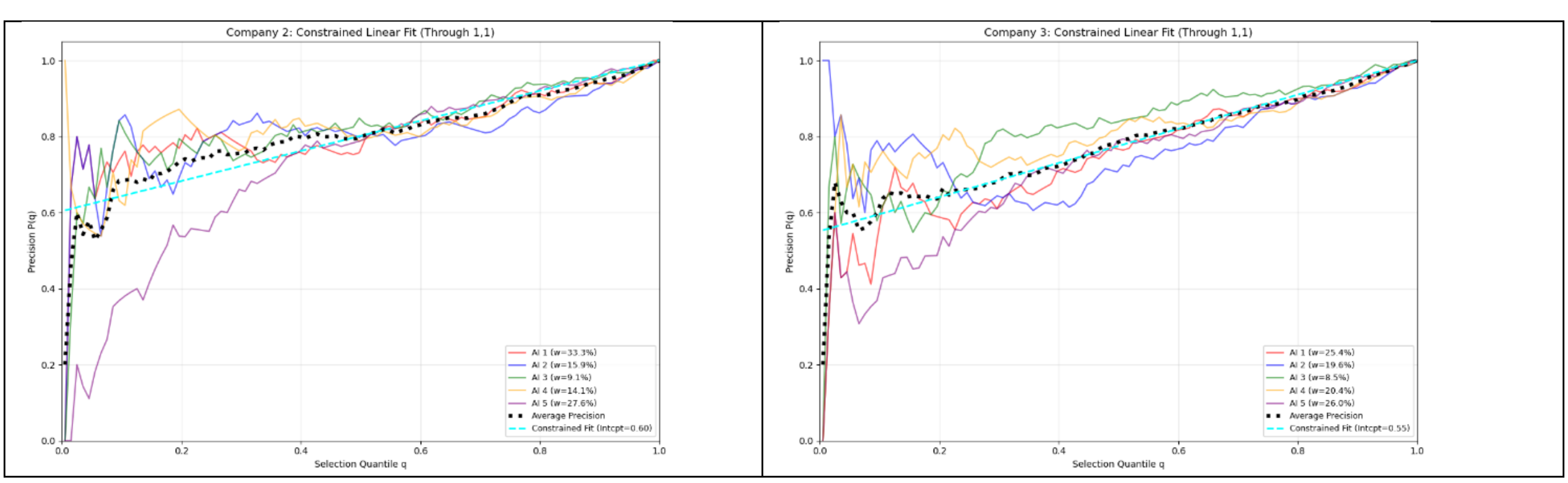

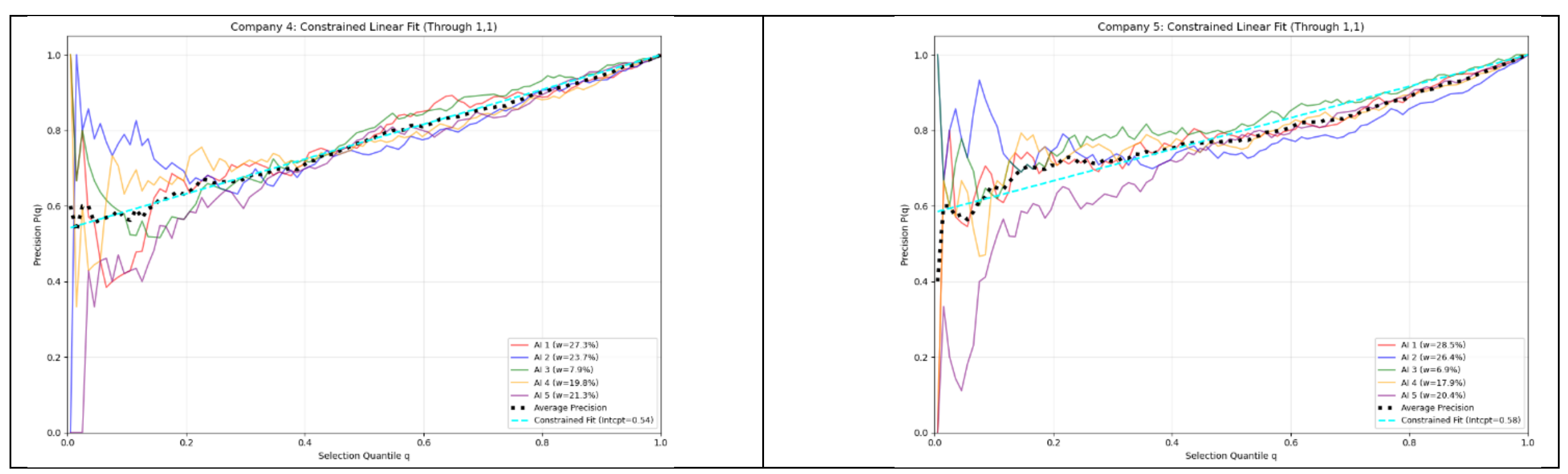

## Comparison of Weighted Rho

We wondered whether the weighted average of the Rhos would do a better job of predicting the intercept. It doesn't

| Company | **1** | **2** | **3** | **4** | **5** |
|---|---|---|---|---|---|
| Average correlation | 0.545 | 0.618 | 0.493 | 0.471 | 0.539 |
| Intercept | 0.591 | 0.604 | 0.551 | 0.540 | 0.583 |
| % Difference | +8.5% | -2.2% | +10.6% | +14.7% | +8.3% |
| Weighted Rho | 0.793 | 0.831 | 0.772 | 0.757 | 0.792 |

## Figure 5 for the other companies

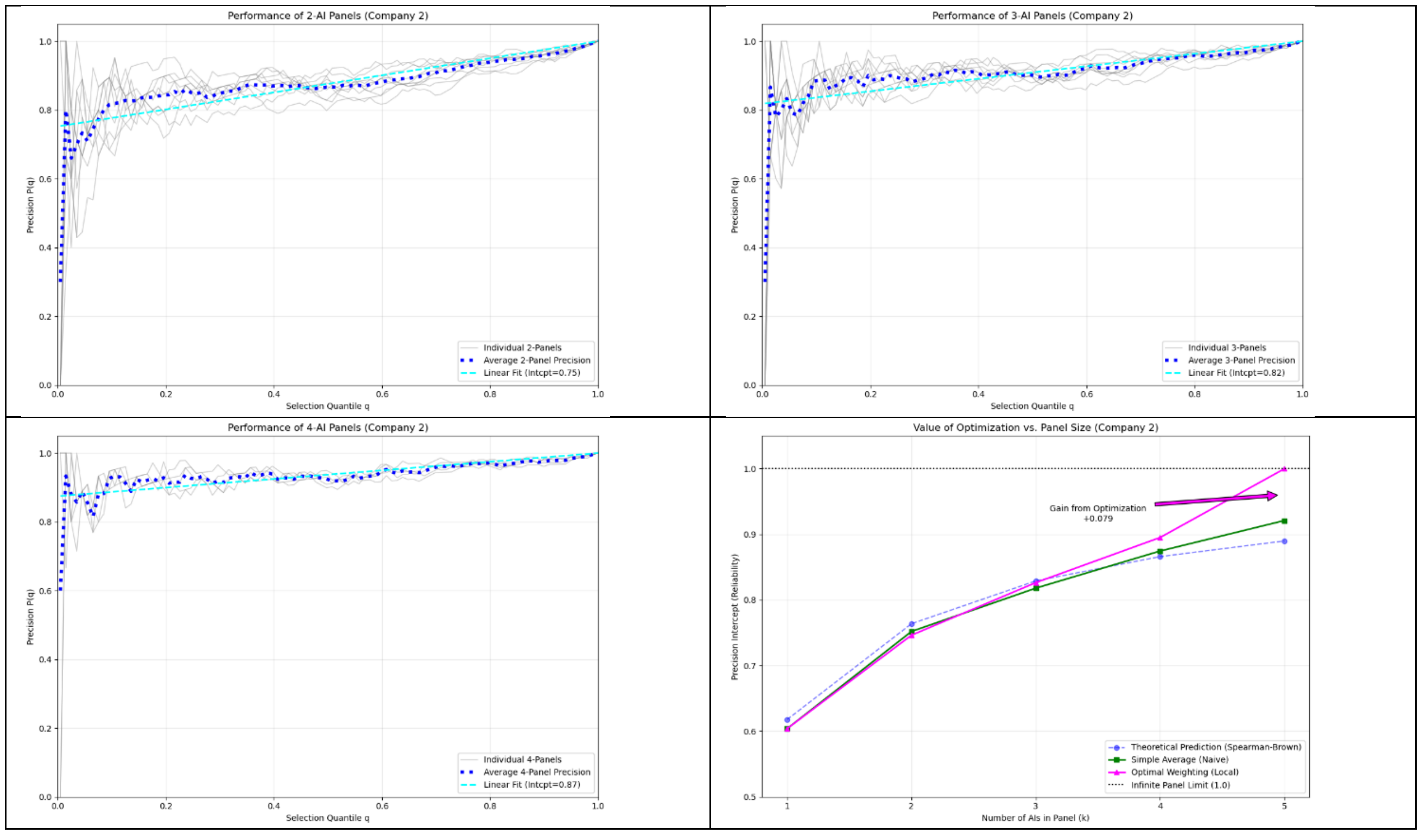

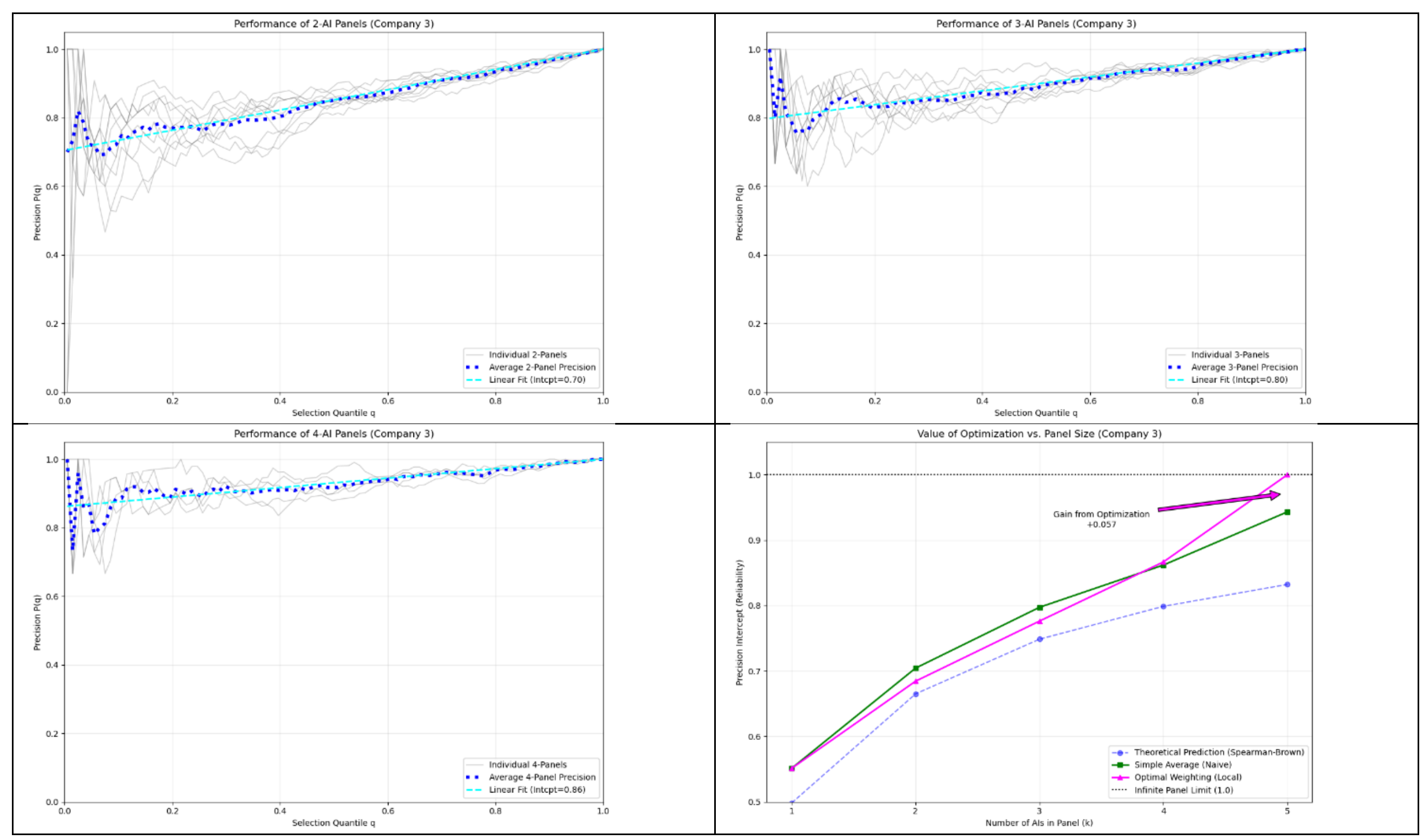

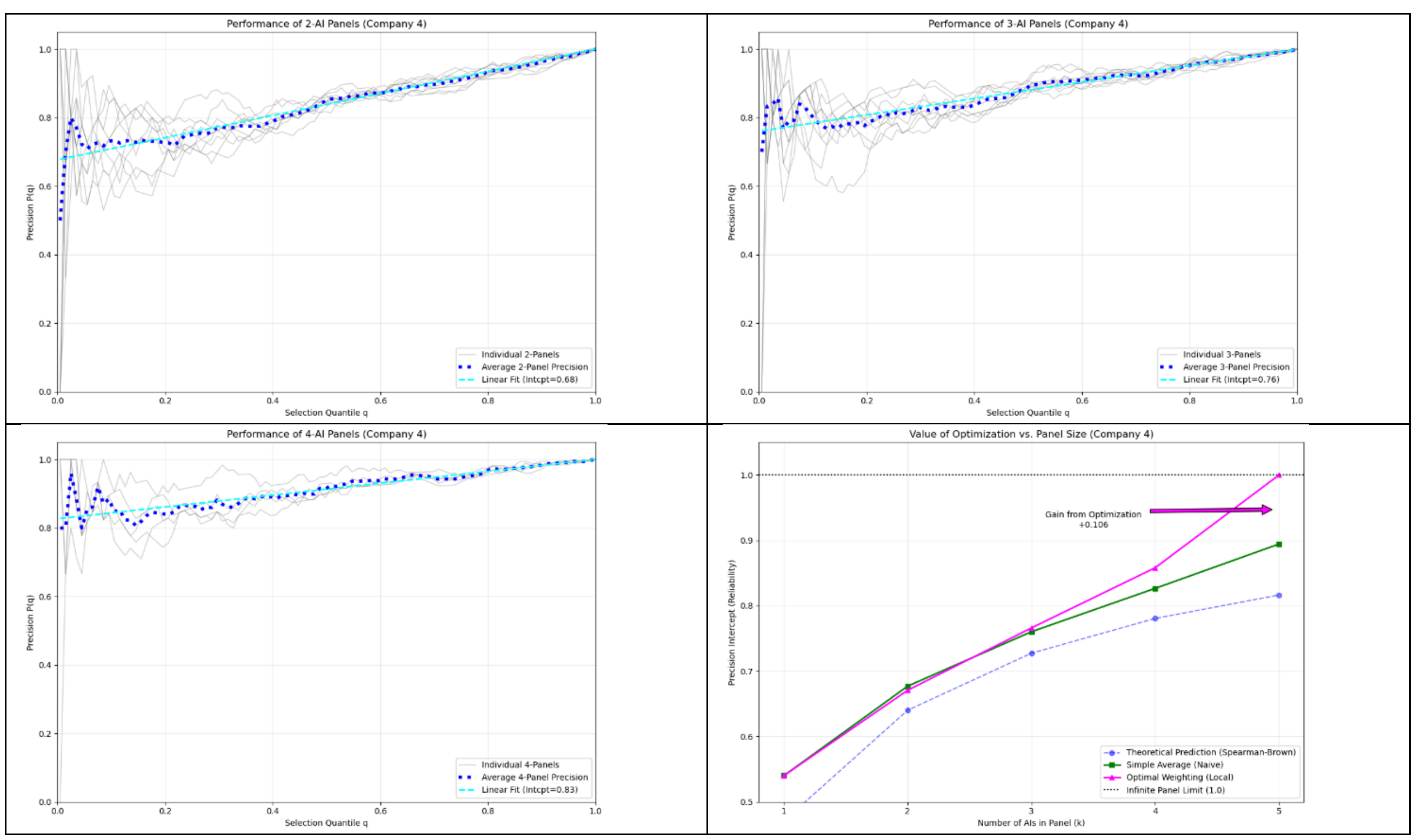

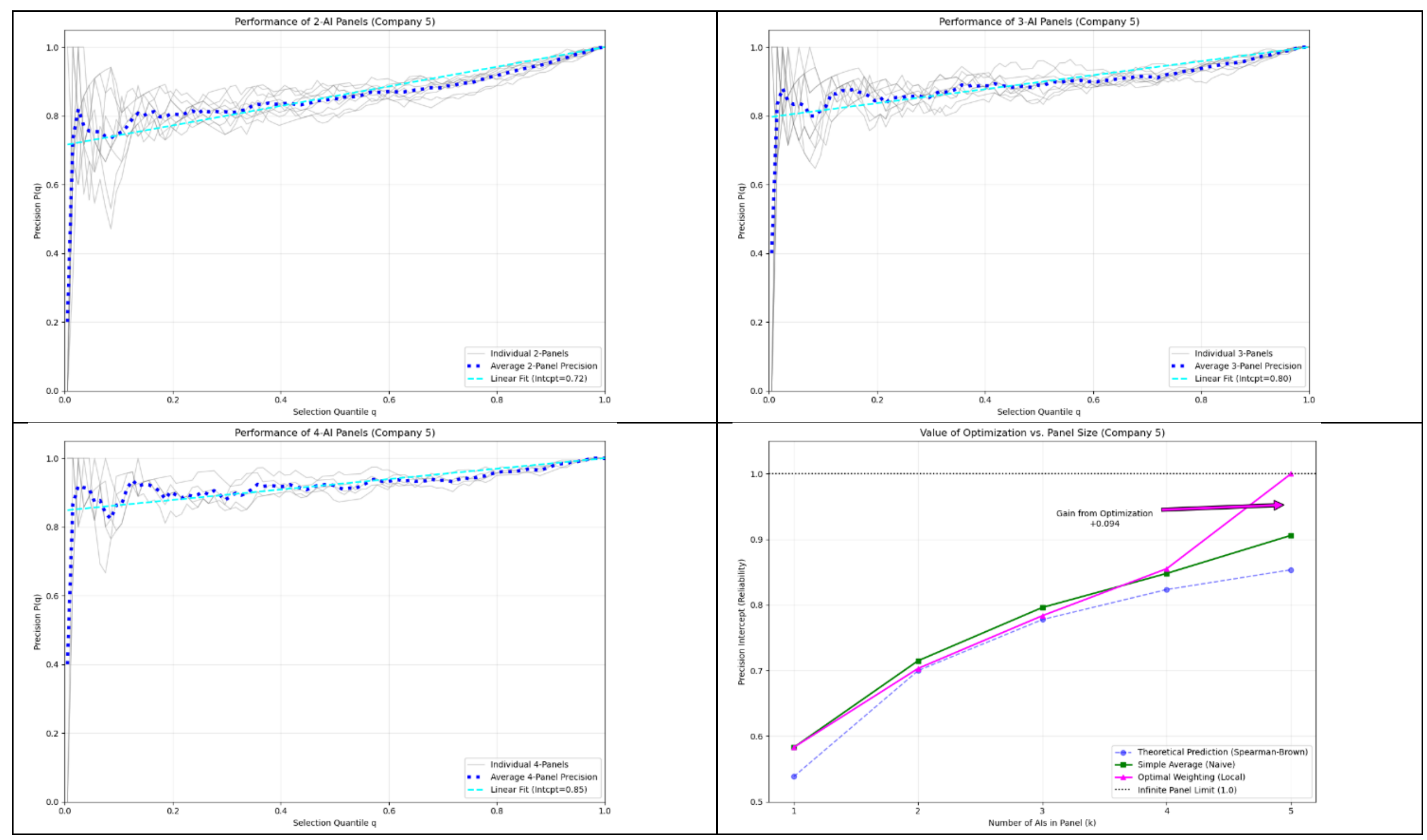

## Observed statistics of correlations (across all companies)

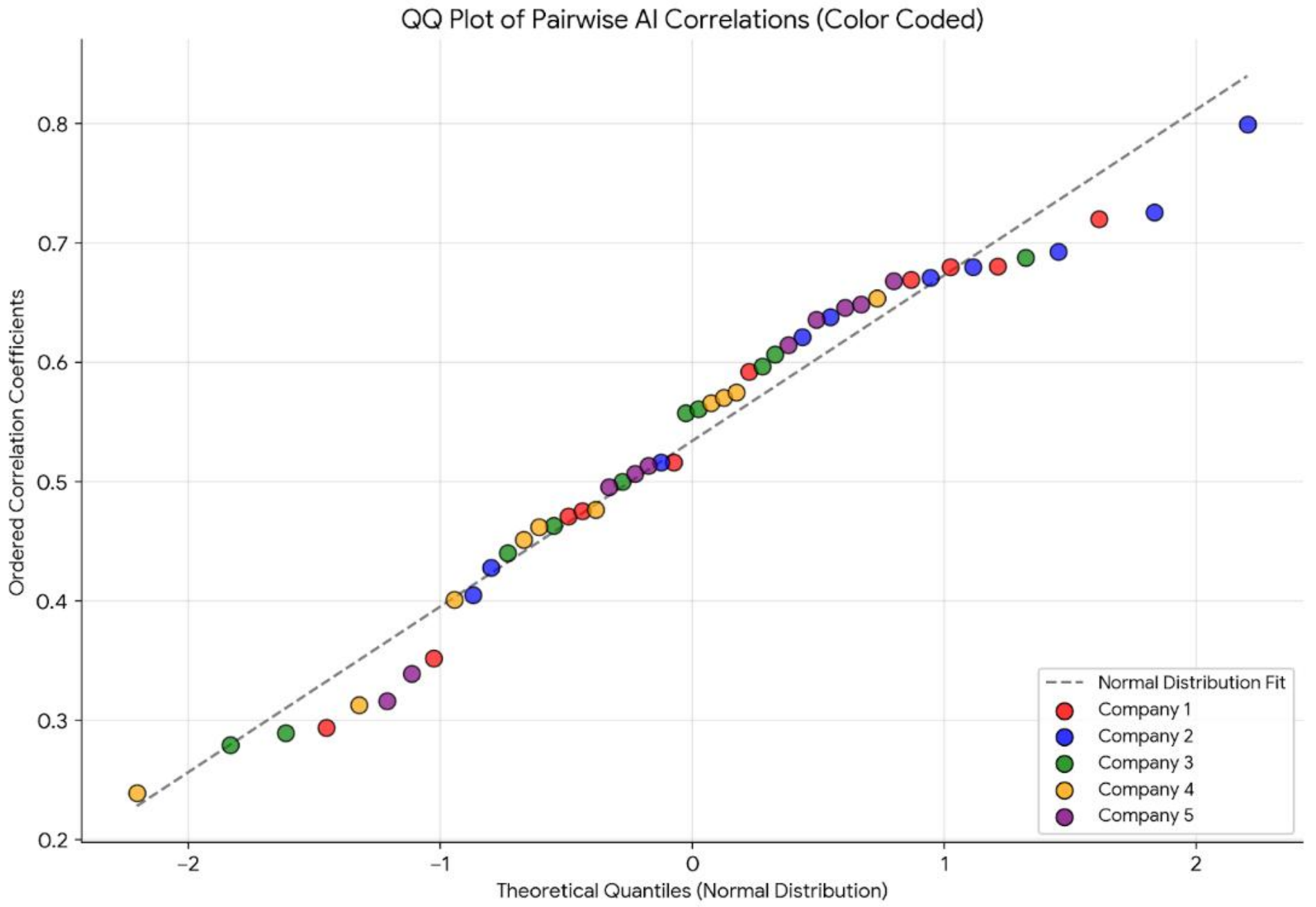

```
--- Correlation Stats per Company ---
Company 1: Mean rho = 0.545
Company 2: Mean rho = 0.618
Company 3: Mean rho = 0.498
Company 4: Mean rho = 0.471
Company 5: Mean rho = 0.539

--- GLOBAL SIMULATION PARAMETERS ---
Grand Mean Correlation (mu): 0.534
Standard Deviation (sigma):  0.136
Min/Max Observed:            0.239 / 0.800
```

## Relationship between accuracy and variance in the data – using Optimal Weights

```
Empirical Variance-Quality Link Check v1.0
------------------------------------------------------------
Company          | AI #   | Variance    | Correlation (Quality)
-----------------------------------------------------------------
Company 1        | 1      | 0.222       | 0.813
Company 1        | 2      | 0.190       | 0.839
Company 1        | 3      | 0.872       | 0.789
Company 1        | 4      | 0.519       | 0.853
Company 1        | 5      | 0.287       | 0.674
Company 2        | 1      | 0.185       | 0.848
Company 2        | 2      | 0.388       | 0.853
Company 2        | 3      | 0.676       | 0.846
Company 2        | 4      | 0.439       | 0.873
Company 2        | 5      | 0.223       | 0.737
Company 3        | 1      | 0.259       | 0.751
Company 3        | 2      | 0.335       | 0.789
Company 3        | 3      | 0.770       | 0.814
Company 3        | 4      | 0.322       | 0.825
Company 3        | 5      | 0.252       | 0.683
Company 4        | 1      | 0.228       | 0.773
Company 4        | 2      | 0.263       | 0.815
Company 4        | 3      | 0.787       | 0.765
Company 4        | 4      | 0.316       | 0.783
Company 4        | 5      | 0.293       | 0.649
Company 5        | 1      | 0.180       | 0.814
Company 5        | 2      | 0.194       | 0.808
Company 5        | 3      | 0.745       | 0.830
Company 5        | 4      | 0.287       | 0.822
Company 5        | 5      | 0.251       | 0.684
-----------------------------------------------------------------

Global Correlation between Variance and Quality: 0.218
P-Value: 0.2948
```

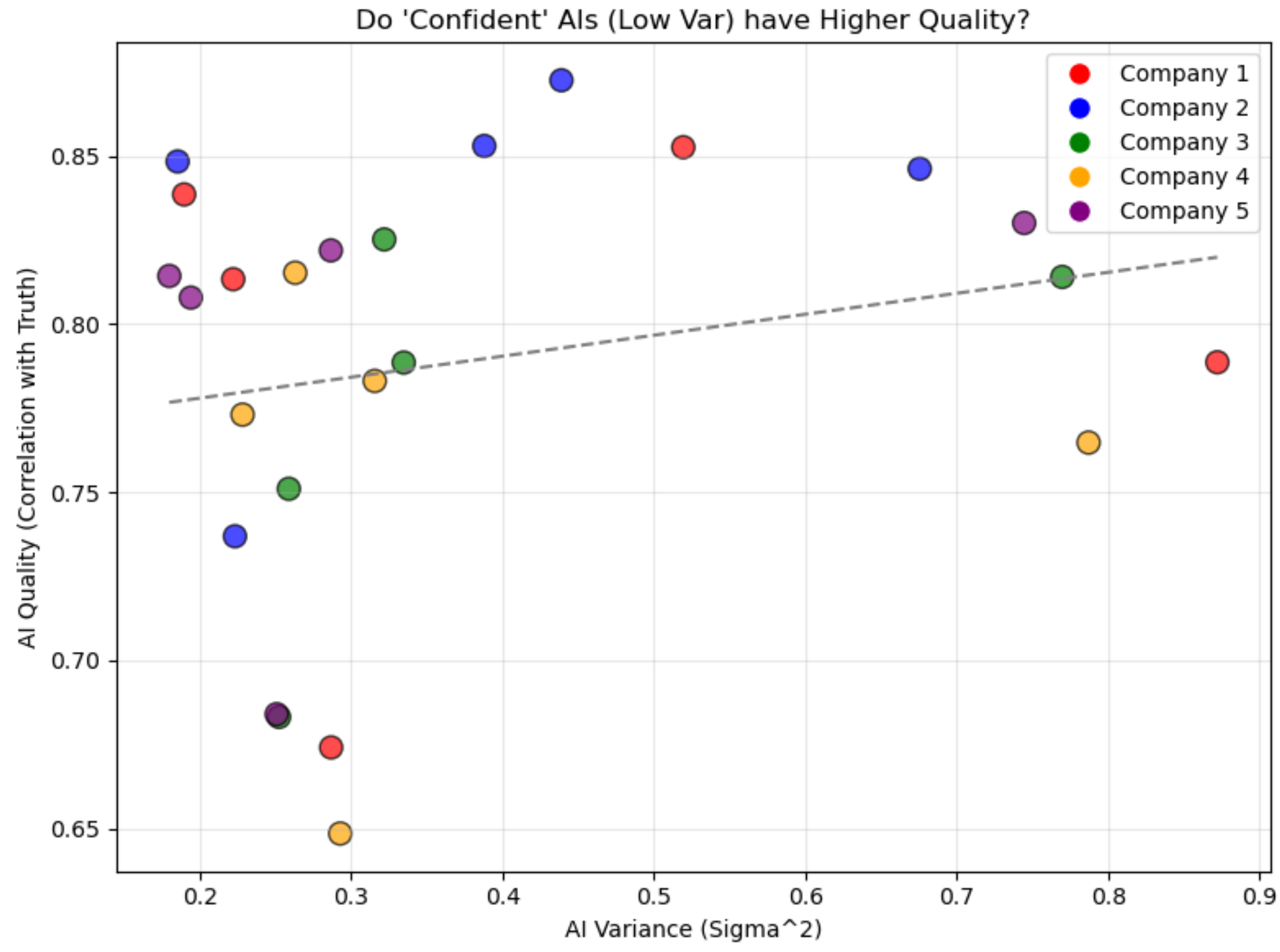

```
Empirical Variance-Quality Check v2.0 (Unweighted Control)
------------------------------------------------------------
Company          | AI #  | Variance    | Correlation (Quality)
-----------------------------------------------------------------
Company 1        | 1     | 0.222       | 0.723
Company 1        | 2     | 0.190       | 0.759
Company 1        | 3     | 0.872       | 0.888
Company 1        | 4     | 0.519       | 0.876
Company 1        | 5     | 0.287       | 0.718
Company 2        | 1     | 0.185       | 0.805
Company 2        | 2     | 0.388       | 0.863
Company 2        | 3     | 0.676       | 0.888
Company 2        | 4     | 0.439       | 0.891
Company 2        | 5     | 0.223       | 0.711
Company 3        | 1     | 0.259       | 0.724
Company 3        | 2     | 0.335       | 0.771
Company 3        | 3     | 0.770       | 0.878
Company 3        | 4     | 0.322       | 0.820
Company 3        | 5     | 0.252       | 0.665
Company 4        | 1     | 0.228       | 0.706
Company 4        | 2     | 0.263       | 0.752
Company 4        | 3     | 0.787       | 0.864
Company 4        | 4     | 0.316       | 0.780
Company 4        | 5     | 0.293       | 0.676
Company 5        | 1     | 0.180       | 0.762
Company 5        | 2     | 0.194       | 0.730
Company 5        | 3     | 0.745       | 0.909
Company 5        | 4     | 0.287       | 0.830
Company 5        | 5     | 0.251       | 0.722
-----------------------------------------------------------------

Global Correlation between Variance and Quality: 0.781
P-Value: 0.0000
```

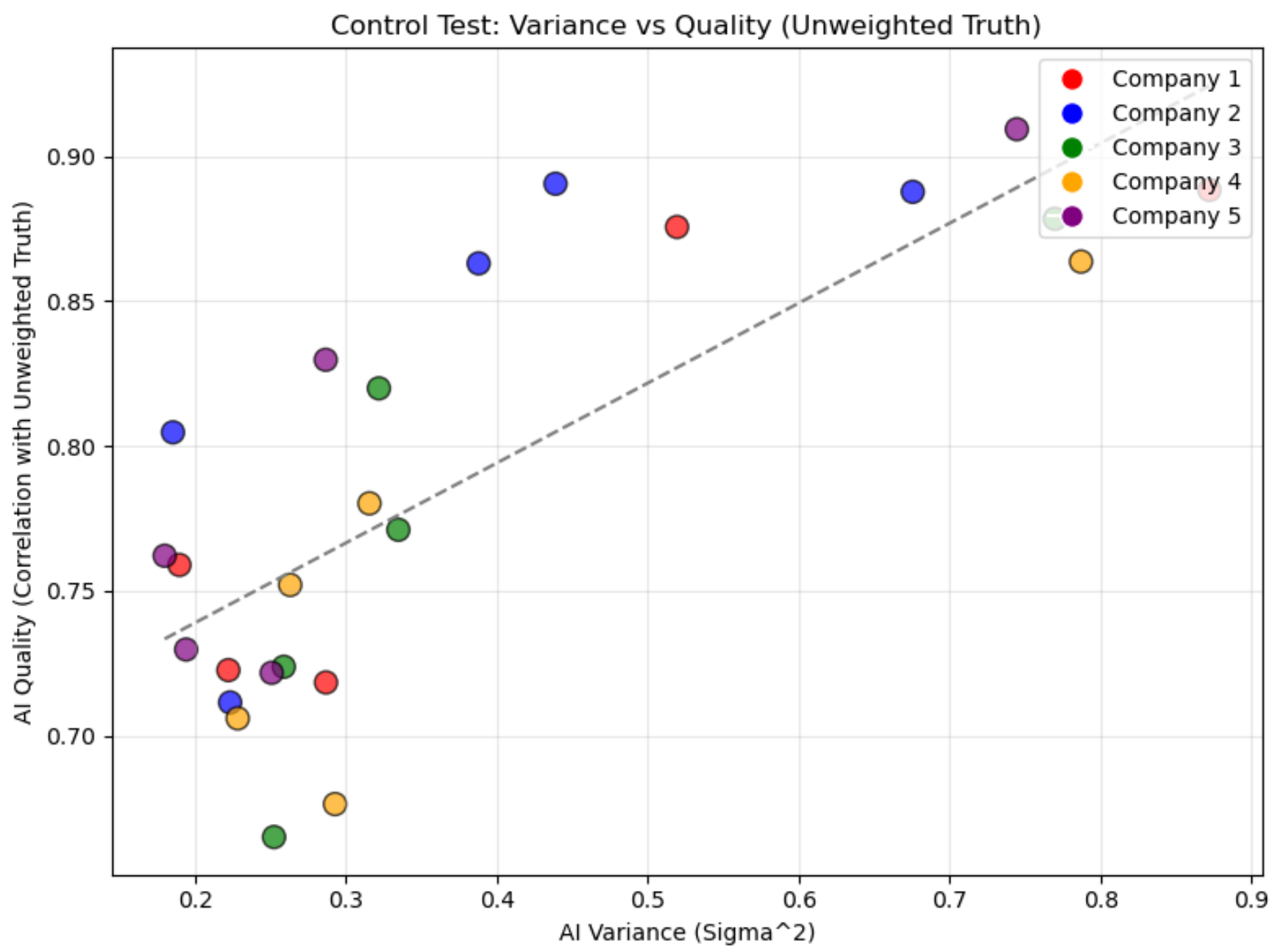

## Exploration of b Curves

```
Done in 20.91 minutes.

--- SMOOTHED RESULTS SUMMARY ---
      q  target_rho  measured_rho    best_b
0  0.05        0.15      0.145658  0.731699
1  0.05        0.20      0.181218  0.716483
2  0.05        0.25      0.245278  0.691827
3  0.05        0.30      0.283723  0.625401
4  0.05        0.35      0.338311  0.594890
5  0.05        0.40      0.382246  0.548849
```

| | | | | |
|---|---|---|---|---|
| 6 | 0.05 | 0.45 | 0.444299 | 0.520547 |
| 7 | 0.05 | 0.50 | 0.466569 | 0.472793 |
| 8 | 0.05 | 0.55 | 0.545163 | 0.508666 |
| 9 | 0.05 | 0.60 | 0.589569 | 0.418930 |
| 10 | 0.05 | 0.65 | 0.639826 | 0.396275 |
| 11 | 0.05 | 0.70 | 0.689425 | 0.430253 |
| 12 | 0.05 | 0.75 | 0.735160 | 0.366456 |
| 13 | 0.05 | 0.80 | 0.790704 | 0.282126 |
| 14 | 0.05 | 0.85 | 0.843252 | 0.259419 |
| 15 | 0.05 | 0.90 | 0.884295 | 0.231366 |
| 16 | 0.05 | 0.95 | 0.940554 | 0.106359 |
| 17 | 0.10 | 0.15 | 0.140770 | 0.830535 |
| 18 | 0.10 | 0.20 | 0.189855 | 0.762016 |
| 19 | 0.10 | 0.25 | 0.235525 | 0.722004 |
| 20 | 0.10 | 0.30 | 0.284440 | 0.712434 |
| 21 | 0.10 | 0.35 | 0.342284 | 0.680379 |
| 22 | 0.10 | 0.40 | 0.393238 | 0.608008 |
| 23 | 0.10 | 0.45 | 0.438406 | 0.608207 |
| 24 | 0.10 | 0.50 | 0.474885 | 0.566742 |
| 25 | 0.10 | 0.55 | 0.518366 | 0.541427 |
| 26 | 0.10 | 0.60 | 0.565306 | 0.513924 |
| 27 | 0.10 | 0.65 | 0.629489 | 0.441954 |
| 28 | 0.10 | 0.70 | 0.702887 | 0.423960 |
| 29 | 0.10 | 0.75 | 0.739505 | 0.365279 |
| 30 | 0.10 | 0.80 | 0.804754 | 0.268144 |
| 31 | 0.10 | 0.85 | 0.824420 | 0.299644 |
| 32 | 0.10 | 0.90 | 0.881650 | 0.222477 |
| 33 | 0.10 | 0.95 | 0.941234 | 0.058848 |
| 34 | 0.15 | 0.15 | 0.143404 | 0.841037 |
| 35 | 0.15 | 0.20 | 0.190022 | 0.815923 |
| 36 | 0.15 | 0.25 | 0.240794 | 0.772396 |
| 37 | 0.15 | 0.30 | 0.286885 | 0.710711 |
| 38 | 0.15 | 0.35 | 0.328636 | 0.694098 |
| 39 | 0.15 | 0.40 | 0.396973 | 0.653655 |
| 40 | 0.15 | 0.45 | 0.430793 | 0.620632 |
| 41 | 0.15 | 0.50 | 0.481142 | 0.569759 |
| 42 | 0.15 | 0.55 | 0.544683 | 0.578720 |
| 43 | 0.15 | 0.60 | 0.572305 | 0.585178 |
| 44 | 0.15 | 0.65 | 0.643953 | 0.508956 |
| 45 | 0.15 | 0.70 | 0.684408 | 0.464370 |
| 46 | 0.15 | 0.75 | 0.736269 | 0.395770 |
| 47 | 0.15 | 0.80 | 0.781861 | 0.385822 |
| 48 | 0.15 | 0.85 | 0.839542 | 0.334675 |
| 49 | 0.15 | 0.90 | 0.888453 | 0.205448 |
| 50 | 0.15 | 0.95 | 0.946281 | 0.133328 |
| 51 | 0.20 | 0.15 | 0.142813 | 0.888744 |
| 52 | 0.20 | 0.20 | 0.178168 | 0.850244 |
| 53 | 0.20 | 0.25 | 0.225008 | 0.802030 |
| 54 | 0.20 | 0.30 | 0.285945 | 0.760526 |
| 55 | 0.20 | 0.35 | 0.336562 | 0.722895 |
| 56 | 0.20 | 0.40 | 0.373753 | 0.691326 |
| 57 | 0.20 | 0.45 | 0.441978 | 0.640847 |
| 58 | 0.20 | 0.50 | 0.465601 | 0.639629 |
| 59 | 0.20 | 0.55 | 0.546143 | 0.589485 |
| 60 | 0.20 | 0.60 | 0.580530 | 0.515558 |
| 61 | 0.20 | 0.65 | 0.628672 | 0.486592 |
| 62 | 0.20 | 0.70 | 0.689190 | 0.494493 |
| 63 | 0.20 | 0.75 | 0.731964 | 0.460032 |
| 64 | 0.20 | 0.80 | 0.787999 | 0.367859 |
| 65 | 0.20 | 0.85 | 0.834046 | 0.351093 |
| 66 | 0.20 | 0.90 | 0.900561 | 0.247383 |
| 67 | 0.20 | 0.95 | 0.946846 | 0.158864 |
| 68 | 0.25 | 0.15 | 0.136603 | 0.923006 |
| 69 | 0.25 | 0.20 | 0.194106 | 0.864982 |
| 70 | 0.25 | 0.25 | 0.229665 | 0.806666 |
| 71 | 0.25 | 0.30 | 0.293656 | 0.785097 |
| 72 | 0.25 | 0.35 | 0.326417 | 0.732165 |
| 73 | 0.25 | 0.40 | 0.378906 | 0.681944 |
| 74 | 0.25 | 0.45 | 0.417125 | 0.678917 |
| 75 | 0.25 | 0.50 | 0.486785 | 0.600828 |
| 76 | 0.25 | 0.55 | 0.536608 | 0.613759 |
| 77 | 0.25 | 0.60 | 0.574084 | 0.562010 |
| 78 | 0.25 | 0.65 | 0.638797 | 0.544943 |
| 79 | 0.25 | 0.70 | 0.701012 | 0.452287 |
| 80 | 0.25 | 0.75 | 0.724851 | 0.451116 |
| 81 | 0.25 | 0.80 | 0.786078 | 0.415780 |
| 82 | 0.25 | 0.85 | 0.838703 | 0.343243 |

```
83   0.25        0.90      0.889485  0.251187
84   0.25        0.95      0.938916  0.237801
85   0.30        0.15      0.139567  0.928652
86   0.30        0.20      0.187508  0.865812
87   0.30        0.25      0.237133  0.828718
88   0.30        0.30      0.282228  0.799233
89   0.30        0.35      0.340333  0.744470
90   0.30        0.40      0.378085  0.731551
91   0.30        0.45      0.430050  0.676535
92   0.30        0.50      0.478566  0.665896
93   0.30        0.55      0.537237  0.616635
94   0.30        0.60      0.563533  0.605906
95   0.30        0.65      0.642584  0.501129
96   0.30        0.70      0.684426  0.534535
97   0.30        0.75      0.734447  0.454283
98   0.30        0.80      0.790905  0.385981
99   0.30        0.85      0.843860  0.367464
100  0.30        0.90      0.899519  0.288015
101  0.30        0.95      0.931892  0.246873
[Parallel(n_jobs=8)]: Done 102 out of 102 | elapsed: 20.9min finished

q     | Slope    | Intercept  | R-squared
---------------------------------------------
0.05  | -0.659   | 0.825      | 0.971
0.10  | -0.774   | 0.931      | 0.988
0.15  | -0.748   | 0.954      | 0.968
0.20  | -0.776   | 0.989      | 0.987
0.25  | -0.795   | 1.012      | 0.983
0.30  | -0.787   | 1.027      | 0.990
```

## Code for Main Simulations

```
import numpy as np
import scipy.stats as stats
import scipy.optimize as optimize
import pandas as pd
from joblib import Parallel, delayed
import time

# ==========================================
# 1. TAIL-SPECIFIC CONFIGURATION
# ==========================================
# We focus strictly on the "Death Zone"
TAIL_SAMPLES = 4000
TAIL_CORES = 8
TAIL_CANDIDATES = 2000
TAIL_UNIVERSE = 100
MAX_TAIL_K = 30

# Use the Standard Superstar Settings (Real World Scenario)
# This ensures we capture the "Robustness" effect if it exists
KINK = 1.6
BOOST = 0  # NB switch this off to get non-boosted data
SHARPNESS = 3.0
T_MEAN = 7.0
MIN_S, MAX_S = 0.2, 1.2
SIG_RHO = 0.05
RHO_SIG_CORR = 0.78

# The "Debug Grid"
# Fine-grained steps in the danger zone
Q_TAIL_VALUES = [0.02, 0.05, 0.10, 0.15, 0.20]
RHO_TAIL_TARGETS = np.linspace(0.80, 0.99, 10)

# ==========================================
# 2. CORE LOGIC (Renamed helper functions to avoid namespace collision?
# actually, Python functions can be reused, but let's be safe and simple)
# ==========================================

def tail_transform(z_scores, kink, boost, sharpness):
    if boost == 0: return z_scores
    scaled_diff = sharpness * (z_scores - kink)
    smooth_excess = (1.0 / sharpness) * np.logaddexp(0, scaled_diff)
```

```python
    return z_scores + boost * smooth_excess

def tail_precision_calc(y_true, y_est, q):
    m = len(y_true)
    k = int(np.round(q * m))
    if k==0: return 0.0
    true_top = set(np.argsort(y_true)[-k:])
    est_top = set(np.argsort(y_est)[-k:])
    return len(true_top.intersection(est_top)) / k

def tail_law_fit(n, rho, c, q):
    # The NB Law formula for fitting
    c = max(c, 0.001)
    n_eff = np.power(n, c)
    return (n_eff * rho + q * (1 - rho)) / (1 + (n_eff - 1) * rho)

def fit_tail_exponent(n_values, p_observed, measured_rho, q):
    def objective(c):
        p_pred = tail_law_fit(n_values, measured_rho, c, q)
        return np.sum((p_observed - p_pred)**2)
    res = optimize.minimize_scalar(objective, bounds=(0.01, 1.5), method='bounded')
    return res.x

def run_tail_simulation(q, target_rho, seed):
    np.random.seed(seed)

    # 1. Generate Universe
    mu_vec = np.array([target_rho, (MIN_S + MAX_S)/2])
    cov_val = RHO_SIG_CORR * SIG_RHO * 0.2
    cov_mat = np.array([[SIG_RHO**2, cov_val], [cov_val, 0.2**2]])

    try:
        params = stats.multivariate_normal.rvs(mean=mu_vec, cov=cov_mat, size=TAIL_UNIVERSE)
    except:
        params = np.random.multivariate_normal(mu_vec, cov_mat, size=TAIL_UNIVERSE)

    Z_common = stats.norm.rvs(size=TAIL_CANDIDATES)
    scores_raw = []

    for i in range(TAIL_UNIVERSE):
        r = np.clip(params[i, 0], 0.01, 0.999) # Allow closer to 1.0
        s = np.clip(params[i, 1], MIN_S, MAX_S)

        loading = np.sqrt(r)
        noise_scale = np.sqrt(1 - r)
        Z_specific = stats.norm.rvs(size=TAIL_CANDIDATES)

        X_norm = loading * Z_common + noise_scale * Z_specific

        # Apply Superstar Transform
        X_skewed = tail_transform(X_norm, KINK, BOOST, SHARPNESS)

        X_skewed_z = (X_skewed - np.mean(X_skewed)) / np.std(X_skewed)
        X_final = (X_skewed_z * s) + T_MEAN
        scores_raw.append(X_final)

    Xt = np.array(scores_raw).T
    y_true = np.mean(Xt, axis=1)

    # Measure Inter-AI Correlation
    corr_matrix = np.corrcoef(Xt.T)
    sum_corr = np.sum(corr_matrix) - TAIL_UNIVERSE
    measured_rho = sum_corr / (TAIL_UNIVERSE * (TAIL_UNIVERSE - 1))

    # 2. Run Panel sizes
    panel_sizes = np.arange(1, MAX_TAIL_K + 1)
    obs_precisions = []

    for k in panel_sizes:
        batch_prec = []
        for _ in range(TAIL_SAMPLES):
            indices = np.random.choice(TAIL_UNIVERSE, k, replace=False)
            est = np.mean(Xt[:, indices], axis=1)
            p = tail_precision_calc(y_true, est, q)
            batch_prec.append(p)
        obs_precisions.append(np.mean(batch_prec))
```

```
    # 3. Fit b (efficiency exponent)
    best_b = fit_tail_exponent(panel_sizes, np.array(obs_precisions), measured_rho, q)

    return {'q': q, 'measured_rho': measured_rho, 'best_b': best_b}

# ==========================================
# 3. EXECUTION
# ==========================================
if __name__ == "__main__":
    tail_tasks = []
    seed_counter = 5000 # distinct seeds
    for q in Q_TAIL_VALUES:
        for rho in RHO_TAIL_TARGETS:
            tail_tasks.append((q, rho, seed_counter))
            seed_counter += 1

    print(f"Starting High-Rho Debug (0.80 - 0.99) on {TAIL_CORES} cores...")

    start_time = time.time()
    tail_results = Parallel(n_jobs=TAIL_CORES, verbose=5)(
        delayed(run_tail_simulation)(q, rho, seed) for q, rho, seed in tail_tasks
    )
    end_time = time.time()
    print(f"\nDone in {(end_time - start_time)/60:.2f} minutes.")

    df_tail = pd.DataFrame(tail_results)
    # Sort for readability
    print(df_tail.sort_values(by=['q', 'measured_rho']).to_string())
    # Save separate file
    df_tail.to_csv("tail_zone_debug.csv", index=False)
```